\documentclass[draftcls,11pt]{IEEEtran}
\onecolumn 
\usepackage[cmex10]{amsmath}
\usepackage{xcolor}
\usepackage[multiple]{footmisc}

\usepackage{mdwtab}
\usepackage{eqparbox}
\usepackage{tabularx}
\usepackage{graphicx,amssymb,rangecite,upgreek,graphicx,dsfont,mathrsfs}
\usepackage{datetime}
\usepackage{algorithm}
\usepackage{algorithmic} 
\usepackage{comment}
\usepackage[usenames,dvipsnames]{pstricks}
 \usepackage{epsfig}

 \usepackage{pst-grad} 
 \usepackage{pst-plot} 
\newcommand {\exe} {\stackrel{\cdot} {=}}

\newcommand {\bh} {\mbox{\boldmath $h$}}

\newcommand {\bxt} {\mbox{\footnotesize\boldmath $x$}}

\newcommand {\bomegat} {\mbox{\footnotesize\boldmath $\omega$}}
\newcommand {\bgammat} {\mbox{\footnotesize\boldmath $\gamma$}}
\newcommand {\byt} {\mbox{\footnotesize\boldmath $y$}}

\newcommand {\bx} {\mbox{\boldmath $x$}}
\newcommand {\by} {\mbox{\boldmath $y$}}

\newcommand {\bA} {\mbox{\boldmath $A$}}

\newcommand {\bE} {\mathbb{E}}

\newcommand {\bX} {\mbox{\boldmath $X$}}
\newcommand {\bY} {\mbox{\boldmath $Y$}}

\newcommand{\calA}{{\cal A}}

\newcommand{\calF}{{\cal F}}

\newcommand{\calH}{{\cal H}}
\newcommand{\calI}{{\cal I}}

\newcommand{\calL}{{\cal L}}

\newcommand{\calP}{{\cal P}}

\newcommand{\calT}{{\cal T}}

\newcommand{\calV}{{\cal V}}
\newcommand{\calW}{{\cal W}}

\def\bgamma{{\mbox{\boldmath $\gamma$}}}

\def\bomega{{\mbox{\boldmath $\omega$}}}

\def\bomega{{\mbox{\boldmath $\omega$}}}

\def\bvarphi{{\mbox{\boldmath $\varphi$}}}

\def\bvarphit{{\mbox{\boldmath $\varphi$}}}

\def\balpha{{\mbox{\boldmath $\alpha$}}}

\newcommand{\be}{\begin{equation}}
\newcommand{\ee}{\end{equation}}
\newcommand{\beqna}{\begin{eqnarray}}
\newcommand{\eeqna}{\end{eqnarray}}

\usepackage{theorem}
\DeclareFontFamily{U}{mathx}{\hyphenchar\font45}
\DeclareFontShape{U}{mathx}{m}{n}{
      <5> <6> <7> <8> <9> <10>
      <10.95> <12> <14.4> <17.28> <20.74> <24.88>
      mathx10
      }{}
\DeclareSymbolFont{mathx}{U}{mathx}{m}{n}

\DeclareMathSymbol{\bigtimes}{1}{mathx}{"91}
\theorembodyfont{\rmfamily}

\newcommand{\Ind}{{\mathds{1}}}

\newcommand{\abs}[1]{\left|#1\right|}

\newcommand{\bOmega}{{\mathbf{\Omega}}}
\newcommand{\barOmega}{{\bar{\mathbf{\Omega}}}}

\DeclareMathOperator{\tr}{tr}
\theoremheaderfont{\itshape}

\newtheorem{theorem}{Theorem}

\newtheorem{lemma}{Lemma}

\newcommand{\p}[1]{\left(#1\right)}
\newcommand{\pp}[1]{\left[#1\right]}
\newcommand{\ppp}[1]{\left\{#1\right\}}
\newcommand{\norm}[1]{\left\|#1\right\|}

\begin{document}

\title{Gaussian Intersymbol Interference Channels With Mismatch}

\author{Wasim~Huleihel ~Salman~Salamatian ~Neri~Merhav ~Muriel~M\'edard
\thanks{\hspace{-0.35cm}W. Huleihel, S. Salamatian, and M. M\'edard are with the Research Laboratory of Electronics at the Massachusetts Institute of Technology, Cambridge, MA (e-mail: \{wasimh,salmansa,medard\}@mit.edu). The work of W. Huleihel was supported by the MIT-Technion Postdoctoral Fellowship.

N. Merhav is with the Andrew \& Erna Viterbi Faculty of Electrical Engineering at the Technion-Israel Institute of Technology, Haifa 3200003, Israel (e-mail: merhav@ee.technion.ac.il).}
}
\maketitle

\IEEEpeerreviewmaketitle

\allowdisplaybreaks

\abstract
This paper considers the problem of channel coding over Gaussian intersymbol interference (ISI) channels with a given metric decoding rule. Specifically, it is assumed that the mismatched decoder has an incorrect assumption on the impulse response function. The mismatch capacity is the highest achievable rate for a given decoding rule. Existing lower bounds to the mismatch capacity for channels and decoding metrics with memory (as in our model) are presented only in the form of multi-letter expressions that have not been calculated in practice. Consequently, they provide little insight on the mismatch problem. In this paper, we derive computable single-letter lower bounds to the mismatch capacity, and discuss some implications of our results. Our achievable rates are based on two ensembles; the ensemble of codewords generated by an autoregressive process, and the ensemble of codewords drawn uniformly over a ``type class" of real-valued sequences. Computation of our achievable rates demonstrates non-trivial behavior of the achievable rates as a function of the mismatched parameters. As a simple application of our technique, we derive also the random coding exponent associated with a mismatched decoder which assumes that there is no ISI at all. Finally, we compare our results with universal decoders which are designed \emph{outside} the true class of channels that we consider in this paper. 

\section{Introduction}

The mismatch capacity is the highest achievable rate for a given, possibly suboptimal, decoding rule. This scenario arises naturally when, due to imprecise channel measurement, the receiver performs maximum-likelihood decoding with respect to the wrong channel law, or when the receiver is intentionally designed to perform a suboptimal decoding rule due to implementation constraints. This problem has been studied extensively, see e.g., \cite{MerhavLapidoth,LapidothNarayan,Ganti,SomekhGene} and many references therein. Finding a single-letter expression for the mismatch capacity is a long-standing open problem.

Most of the existing work on the mismatch capacity has focused on deriving achievable rates using random coding arguments for memoryless channels and decoding metrics. For a given block length, one typically selects a certain ensemble of rate--$R$ codes and then studies the highest achievable rate for which the average probability of error still tends to zero as the 
block length tends to infinity. Different random coding ensembles yield different lower bounds to the mismatch capacity. For example, the ensemble of identically and independently distributed (i.i.d.) codewords leads to the generalized mutual information (GMI) rate \cite{Stiglitz,KaplanSh,abou_faycal}. Tighter lower bounds to the mismatch capacity can be derived using constant-composition ensembles \cite{Graph_decomposition,Hui}, and cost-constrained ensembles \cite{Ganti,ScarlettMis}. Although the GMI is the weakest bound of this class, it has the advantage of being applicable also to channels over infinite alphabets, as its derivation relies on Gallager's bounding technique \cite{Gallager} rather than on the method of types. While superior to the GMI, the bound based on the constant-composition ensemble, relies heavily on the method of types\footnote{More critically, the method of types is of limited applicability to channels with memory, rendering the bound inapplicable to such channels.} \cite{CsisKro} and thus, at least at first glance, limited to channels over finite alphabets. See, however, \cite{MerhavLapidoth} for some extensions to memoryless channels of an exponential type and to some channels with memory. In \cite{Ganti}, this bound was also extended to general alphabets using an alternative derivation that does not require the method of types. In \cite{AbbeMedard}, the question of finding the best mismatched decoder (in the sense of maximizing the achievable rate) over a given family of linear decoders was considered, along with an efficient algorithm for computing this decoder. Finally, \cite{ScarlettMis} considered a more comprehensive analysis of the random-coding error probability under various ensembles, including error exponents, second-order coding rates \cite{Polyanskiy,Hayashi}, and refined asymptotic results based on the saddlepoint approximation \cite{Jensen}. In the discrete case, the results of \cite{ScarlettMis} are tight in the error exponent sense, but for general alphabets there is no guarantee for ensemble tightness, as the analysis is based on Gallager's bounding technique.

For channels and decoding rules with memory, however, there are no known single-letter lower bounds, even in specific examples. The only existing lower bound, derived in \cite{Ganti}, which holds for a general family of channels and decoding metrics with memory, appears in the form of a multi-letter expression. Unfortunately, this expression cannot be calculated in practice and it provides only little insight on the mismatch decoding problem. 

Motivated by the last paragraph, in this work, we consider a specific class of channels with memory; Gaussian intersymbol interference (ISI) channels, with a mismatched decoding metric that is based upon wrong ISI coefficients (see Section~\ref{sec:problemFormulation} for a precise definition of our model). Considering this problem is important when, for example, the depth of the ISI is large (i.e, many taps), and thus the implementation of the optimal maximum-likelihood (ML) decoder is complicated. In such cases, one might want to intentionally limit the depth of the assumed ISI in decoding metric so that to keep the decoding complexity within reasonable limits. Another possible motivation is that when the channel is slowly time-varying (e.g., fading), it might limit the block length one can work with (block length within which the channel is nearly fixed), and then estimation errors resulting from channel estimation can be significant. In such cases it is interesting to understand how this issue affects the achievable rates. As was demonstrated in \cite{NarayanCsisz}, even for discrete memoryless channels (DMCs) and memoryless decoding metrics, the ensemble of i.i.d. input codewords is not optimal, and an improved bound on the mismatch capacity of the DMC can be obtained through a random coding argument applied to a superalphabet, or equivalently, inputs defined over product spaces (i.e., inputs with memory). 

We consider two random coding ensembles. In the first, the random codewords are generated by an autoregressive (see eq. \eqref{ARmodel} for more details). For this ensemble, we derive a simple and computable single-letter lower bound to the mismatch capacity. The obtained rate is ensemble-tight, namely, it captures the exact maximum achievable rate for which the ensemble average error probability vanishes. Also, contrary to the above-mentioned multi-letter expressions, in the Gaussian ISI case, our achievable-rate formula is given in terms of frequency-domain integrals of certain spectral quantities, which are computable at least numerically. The main technical contribution in the derivation is a novel procedure to assess the exponential behavior of the error probability, using the saddle-point integration method (see, e.g., \cite{Bruijn}). Specifically, as shall be seen in the proof of our main results, the probability of error associated with our mismatched decoder can be written as a function of the volumes (Lebesgue measure) of some ``conditional typical set" of sequences with continuous-valued components. This typical set, of some input sequence $(x_1,x_2,\ldots,x_n)$, given an output sequence $(y_1,y_2,\ldots,y_n)$, will contain all sequences which, within $\epsilon>0$, have the same sufficient statistics as $(x_1,x_2,\ldots,x_n)$ induced by our mismatched decoding rule. Accordingly, to analyze the probability of error we need to analyze the volume of this typical set. While this was also the main difficulty of \cite{NeriUni,Wasim3}, and resolved using the ``backward channel" technique, here, we use the saddle-point integration method which is, more direct, and simplifies the derivations significantly. Since we deal with a Gaussian channel, the above mentioned typical set depends on the input sequence only through certain simple statistics, such as, the correlation with the output sequence and auto-correlations (up to some order), and thus it is possible to get ``single-letter" expressions, as opposed to general channels with (finite) memory, where only multi-letter formulas are available.
 
\begin{figure}[!t]
\begin{minipage}[b]{1.0\linewidth}
\centering
\centerline{\includegraphics[width=9cm,height = 7cm]{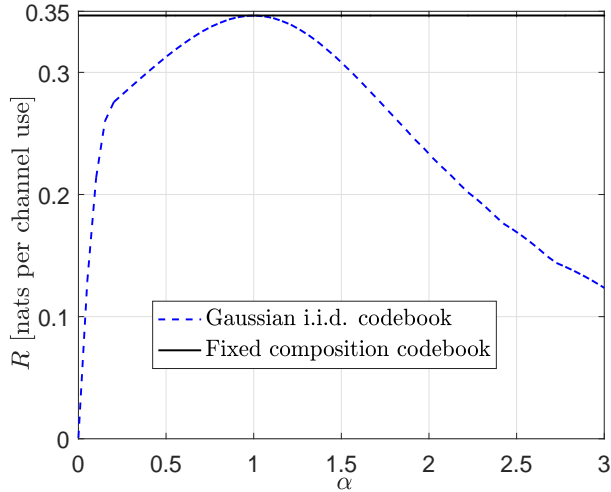}}
\end{minipage}
\caption{Achievable rate as a function of the mismatched level $\alpha$, for the additive white Gaussian noise channel (AWGN), using Gaussian i.i.d. codebook and fixed composition ensemble (i.e., codewords are drawn uniformly at random over the $n$-dimensional hypersphere of radius $\sqrt{nP_X}$), where $P_X=1$ and $\sigma^2=1$. The fixed composition ensemble achieve the capacity of the AWGN, i.e., $1/2\log2$, regardless of the value of $\alpha>0$.}
\label{fig:1}
\end{figure}

Then, using the same methods, we analyze also the ensemble of codewords which drawn according to the uniform distribution within a ``type class" of real-valued sequences \cite{NeriUni,Wasim3} (see eq. \eqref{condtypemar} for more details). As before, for this ensemble we derive a computable single-letter lower bound to the mismatch capacity. The resulting formula is more complicated to compute compared to the previous ensemble. However, the fixed composition ensemble can be better than the autoregressive ensemble. As an illustrating example, consider the simple case where both the true channel and the decoding metric are memoryless. Specifically, the channel is given by $y_t = x_t+w_t$, for $t=1,2,\ldots,n$, where $\ppp{w_t}$ is a white Gaussian noise, while the mismatched decoder computes a ML estimate which correspond to the channel $\bar{y}_t = \alpha x_t+w_t$, for $t=1,2,\ldots,n$, and $\alpha>0$ designates the mismatched parameter. It should be clear then that codewords that are drawn on the entire hypersurface of radius $\sqrt{nP_X}$ achieve the matched capacity irrespectively of the value of $\alpha$.\footnote{This can be easily seen by expanding the mismatched decoding rule along with the fact that the different codewords have the same energy, and comparing to the maximum-liklihood decoder.} This is no longer true if one generates codewords from the autoregressive codebook (in particular, i.i.d. codewords cannot achieve capacity if there is a mismatch, even though the true channel and the decoding metric are memoryless). This is illustrated in Fig.~\ref{fig:1}. In general scenarios (e.g., Fig.~\ref{fig:2}), we found that there is no special order between the autoregressive codebook and the above ``fixed composition" ensemble, in terms of the achievable rates, namely, no ensemble is uniformly better than the other. Nevertheless, it seems that in most cases the fixed composition ensemble is more powerful as a function of the mismatched parameters.

It turns out that a byproduct of our analysis is an ensemble-tight characterization of the random coding error exponent. Exponentially tight analysis of the average probability of error was extensively studied before (see, e.g., \cite{Merhavcodes,MerhavSlep,Merhavoptimalbin,WasimInter}) mainly for discrete memoryless sources and channels. Here, on the other hand, as we deal with sources and channels with memory defined over infinite alphabets, the same methods cannot be applied. Specifically, to assess the exact exponential rate of the average error probability, we need to evaluate the log-volumes of some conditional typical sets of sequences with continuous-valued components. While this was also the main core of \cite{NeriUni,Wasim3}, here, the saddle-point integration method simplifies the analysis considerably. Accordingly, to demonstrate the usefulness of our techniques, we consider the ensemble of codewords drawn on the entire hypersurface of radius $\sqrt{nP_X}$, and derive the \emph{exact} random coding error exponent in the case of a memoryless decoding metric. 

Finally, we consider also the problem of universal decoding which received very much attention in the last four decades \cite{CsisKro,NeriUni,Wasim3,Csis2,ZivUni,LapZiv,FerderLapidoth,merFeder,Lomnitz,Lomnitz2,Misra,Shayevitz,Shayevitz2,Shayevitz3,UniNeri2}. Indeed, as in the mismatch decoding problem, in many practical situations encountered in coded communication systems, the specific channel over which transmission is to be carried out is unknown to the receiver. The receiver only knows that the channel belongs to a given family of channels. In such a case, the implementation of the optimum ML decoder is of course precluded, and thus, universal decoders, independent of the unknown channel, and which perform asymptotically as well as the ML decoder had the channel law been known, are sought. In this paper, we look at the following scenario. Consider the Gaussian ISI channel, and assume that due to complexity issues concerning the implementation of the optimal ML decoder, the receiver intentionally uses a mismatched decoder which corresponds to a memoryless channel (namely, without ISI). Nonetheless, we allow the receiver to optimize his memoryless metric, namely, it can be a function of the true channel. Now, consider a different receiver which uses a universal decoder which is designed for a memoryless channel. In other words, the true family of channels is \emph{outside} the class of channels for which the universal decoder is actually designed, i.e, mismatched universal decoder. Then, which approach yields higher rates? We show that both decoders achieve the same rates, and in fact achieve also the same error exponent. This means that, at least in the specific scenario described above, our results provide indications that universal decoders exhibit a robustness property with respect to (w.r.t.) the family of channels over which they are actually designed. In other words, this observation (potentially) suggests a certain expansion of the classic notion of universality to cases where the true underlying channel is outside the class.

The paper is organized as follows. In Section~\ref{sec:notation} we establish some notation. Then, in Section~\ref{sec:problemFormulation}, we present our system model and formalize the problem. In Section~\ref{sec:mainresults} we assert our main results. Specifically, we first provide achievable rates under the autoregressive random coding ensemble and the fixed composition ensemble, respectively. Then, we consider the problem of mismatch universal decoding. Section~\ref{sec:proof} is devoted to the proofs of our main results. Finally, our conclusions appear in Section~\ref{sec:conc}.
\allowdisplaybreaks

\section{Notation Conventions}\label{sec:notation}
Throughout this paper, scalar random variables (RV's) will be denoted by capital letters, their sample values will be denoted by the respective lower case letters and their alphabets will be denoted by the respective calligraphic letters. A similar convention will apply to random vectors and matrices and their sample values, which will be denoted with same symbols in the bold face font. The expectation operator of a RV $X$ will be denoted by $\bE(X)$. When using vectors and matrices in a linear-algebraic format, $n$-dimensional vectors, like $\bx$, will be understood as column vectors, the operators $\p{\cdot}^T$ and $\p{\cdot}^H$ will denote vector or matrix transposition and vector or matrix conjugate transposition, respectively, and so, $\bX^T$ would be a row vector. The $\ell_2$-norm of a vector $\bx$ is denoted by $\norm{\bx}_2$. For two positive sequences $\ppp{a_n}$ and $\ppp{b_n}$, the notation $a_n\exe b_n$ means equivalence in the exponential order, i.e., $\lim_{n\to\infty}\frac{1}{n}\log\p{a_n/b_n} = 0$, where in this paper, logarithms are defined w.r.t. the natural basis, that is, $\log(\cdot) = \ln(\cdot)$. Given two real numbers $a$ and $b$, we denote by $\pp{a:b}$ the set of integers $\ppp{n\in\mathbb{N}:\;\left\lceil a\right\rceil\leq n\leq \left\lceil b\right\rceil}$, and we let $a\wedge b\triangleq\min(a,b)$ and $a\vee b\triangleq\max(a,b)$. We define $\mathbb{R}_+\triangleq\ppp{x\in\mathbb{R}:\;x>0}$, and $\mathrm{sgn}(\cdot)$ is the sign function, i.e., $\mathrm{sgn}(x) = \frac{\mathrm{d}}{\mathrm{d}x}\abs{x}$, for $x\neq0$. The volume of a set $\calA\subset\mathbb{R}^n$ is defined as $\mathrm{Vol}\ppp{\calA}\triangleq\int_{\calA}\mathrm{d}\bx$. Finally, the indicator function on a set $\calA$ will be denoted by $\Ind\ppp{\calA}$. 

\section{Problem Setting}\label{sec:problemFormulation}
Consider a discrete time, $K$-tap Gaussian ISI channel, characterized by
\begin{align}
y_t = \sum_{i=0}^{K}h_ix_{t-i}+w_t,\label{2tapISI}
\end{align}
for $t=[1:n]$, where $\ppp{x_t}_{t=1-K}^n\in\mathbb{R}^n$ are the channel inputs, subjected to an average power constraint $\sum_{t=1}^n \mathbb{E}X_t^2\leq nP_X$, $\ppp{h_i}_{i=0}^K\in\mathbb{R}^{K+1}$ are the ISI coefficients, 
$\ppp{w_t}_{t=1}^n$ is a zero-mean Gaussian white noise with variance $\sigma^2$, and $\ppp{y_t}_{t=1}^n\in\mathbb{R}^n$ are the channel outputs. It is assumed that $\ppp{w_t}_{t=1}^n$ is statistically independent of $\ppp{x_t}_{t=1}^n$. 
We denote by $W(\by\vert\bx)$, the conditional density of the channel output induced by \eqref{2tapISI}, where $\bx\triangleq(x_{1-K}, \ldots,x_n)$ and $\by\triangleq(y_1,\ldots,y_n)$, and without loss of generality we assume that $x_{1-K}=\cdots=x_{0}=0$, namely, an overhead of zeroes at the beginning of each block. Alternatively, we may assume that $x_{-k} = x_{n-k}$, for $k = 0,\ldots,K-1$, that is, a circularity assumption on the input sequence \cite{HIRTMASSEY}. As long as $K$ is fixed and $n\to\infty$ these assumptions have no influence on either the achievable error exponents or the achievable rates. Accordingly, for any $\bx,\by\in\mathbb{R}^n$,
\begin{align}
\log W(\by\vert\bx) =-\frac{n}{2}\log(2\pi \sigma^2) -\frac{1}{2 \sigma^2}\sum_{t=1}^n\p{y_t-\sum_{i=0}^{K}h_ix_{t-i}}^2.\label{2tapISISpe}
\end{align}
A rate $R$ block code of size $n$ is a set of $M = e^{nR}$ equiprobable $n$-dimensional vectors (codewords), $\bx_i = (x_{i,1},\ldots,x_{i,n})\in\mathbb{R}^n$, for $1\leq i\leq M$, to be transmitted over the channel \eqref{2tapISI}. The decoder, upon receiving $\by\in\mathbb{R}^n$, estimates the message $i$ of the transmitted codeword as the one that maximizes $\log V(\by\vert\bx_i)$, henceforth referred as the \emph{decoding metric}. If this decoding metric is not equivalent to that of the ML decoder \eqref{2tapISISpe}, then we say that the decoder is \emph{mismatched}. 

We assume that the metric $V(\cdot\vert\cdot)$ is equivalent to that of the ML decoder of a Gaussian ISI channel with (possibly) different coefficients. Specifically, for any $\bx,\by\in\mathbb{R}^n$, $V(\by\vert\bx)$ is defined as,\footnote{Without loss of generality, we assume that the length of the mismatched filter is the same as the length of the true filter, with the understanding the one can always tap the shorter coefficients sequence with zeros.}
\begin{align}
\log V(\by\vert\bx) =-\frac{n}{2}\log(2\pi \sigma^2) -\frac{1}{2 \sigma^2}\sum_{t=1}^n\p{y_t-\sum_{i=0}^{K}\alpha_ix_{t-i}}^2\label{VGaussian}
\end{align}
where $\ppp{\alpha_i}_{i=0}^K\in\mathbb{R}^{K+1}$ are the mismatched ISI coefficients. In particular, when $\alpha_i = 0$ for $i =1, \ldots, K$, then the decoder assumes that there is no ISI at all, i.e., the mismatched decoder is equivalent to the optimal ML decoder associated with the additive white Gaussian noise channel (AWGN), namely, $y_t = \alpha_0x_t+w_t$, for $t=[1:n]$. 

An error is said to have occurred if the estimated index $\hat{i}$ differs from the correct one, $i$. A rate $R$ is said to be achievable if, for every $\delta>0$, there exists a sequence of codes $\{\mathfrak{C}_n\}_{n\ge 1}$ indexed by the block length $n$, with $M\geq e^{n(R-\delta)}$ and vanishing error probability $P_{e}(\mathfrak{C}_n)$ when decoding with the metric $V(\cdot\vert\cdot)$. The mismatch capacity, $C_{\mathrm{ISI}}^{\mathrm{Mis}}$ is the supremum of all achievable rates. The goal of this paper is to derive lower bounds to the mismatched capacity $C_{\mathrm{ISI}}^{\mathrm{Mis}}$. 

\section{Main Results}\label{sec:mainresults}

In this section, we present and discuss our main results. Specifically, in Subsection~\ref{subsec:misrates} we present achievable rates for the mismatch decoding problem. We start with the autoregressive ensemble, where codewords are drawn from an autoregressive process, and then we move forward to analyzing the fixed composition ensemble. Following these results, in Subsection~\ref{subsec:misdec}, we consider the problem of mismatched universal decoding, as described in the Introduction.

\subsection{Mismatched Achievable Rates}\label{subsec:misrates}

We establish first some notation. Let $p$ be a non-negative integer, define $\calP$ as the set of all vectors $\bvarphi=(\varphi_1,\ldots,\varphi_p)\in \mathbb{R}^p$, such that all roots of the polynomial $z^{p}-\sum_{i=1}^{p}\varphi_{i}z^{p-i}$ lie strictly within the unit circle, and let $\varphi_0 = 0$. We let $A(\nu)$, for $\nu\in[0,2\pi]$, be the Fourier transform of the sequence $\{ \alpha_k\}_{k=0}^K$, i.e.,
\begin{align}
A(\nu) \triangleq \sum_{k=0}^K \alpha_k e^{-jk \nu},~~~\nu\in[0,2\pi]
\end{align}
where $j\triangleq\sqrt{-1}$. Similarly, $\Phi(\cdot)$ and $H(\cdot)$ are the Fourier transforms of $\ppp{\varphi_k}_{k=0}^p$ and $\ppp{h_k}_{k=0}^K$, respectively. Next, define
\begin{align}
\eta^2\triangleq P_X\cdot\pp{\frac{1}{2\pi}\int_0^{2\pi}\frac{\mathrm{d}\nu}{\abs{1-\Phi(\nu)}^2}}^{-1},\label{calculatenoisevar}
\end{align}
and note that $\eta^2$ depends on the choice of $\bvarphi$. Let $\ppp{\gamma_m}_{m=0}^p$, with $\gamma_{-m}=\gamma_m$, for $m\in\pp{1:p}$, be defined as:
\begin{align}
\gamma_m = \sum_{k=1}^p\varphi_k\gamma_{m-k}+\eta^2\delta_{m},\label{YuleWalker}
\end{align}
where $\delta_{m}$ is the Kronecker delta function. For $\nu\in\left[0,2\pi\right]$, define:
\begin{align}
S_X(\nu)&\triangleq\frac{\eta^2}{\abs{1-\Phi(\nu)}^2},\label{eq:S_X}\\
S_Y(\nu)&\triangleq \abs{ H(\nu)}^2\cdot S_X(\nu)+\sigma^2,
\end{align}
namely, the input and output spectra, respectively, and for $\omega\in\mathbb{R}_+$, 
\begin{align}
f_{\omega}(\nu) &\triangleq \frac{\omega}{2}|A(\nu)|^2+\frac{1}{2\eta^2}\left[1+ |\Phi(\nu)|^2 - 2\cdot\mathrm{Re}(\Phi(\nu))\right].\label{fomega}
\end{align}
Finally, define
\begin{align}
\label{I_bar}
\bar{I}_1(\balpha,\bvarphi) \triangleq& \frac{1}{2}\log(2\eta^2)-\min_{\omega\in\mathbb{R}_+}\left\{\vphantom{\sum_{l=0}^J\sum_{i=0}^K\alpha_lh_i\gamma_{l-i}-\p{\frac{1}{2}\norm{\balpha}_2^2\gamma_0+\sum_{l=1}^J \sum_{k=0}^{J-l}\alpha_k\alpha_{k+l}\gamma_l}}-\frac{1}{4\pi}\int_0^{2\pi}\mathrm{d}\nu\log f_\omega(\nu)+\right.
 \left. \frac{\omega^2}{8\pi}\int_0^{2\pi}\mathrm{d}\nu\frac{S_Y(\nu)\abs{A(\nu)}^2}{f_{\omega}(\nu)}\right. \nonumber\\
& \hspace{2.5cm} \left.-\omega\pp{\sum_{l=0}^K\sum_{i=0}^K\alpha_lh_i\gamma_{l-i}-\p{\frac{1}{2}\norm{\balpha}_2^2\gamma_0+\sum_{l=1}^K \sum_{k=0}^{K-l}\alpha_k\alpha_{k+l}\gamma_l}}\right\}
\end{align}
where in case that $p<K$, $\gamma_k$ for $k=p+1,\ldots,K$, are calculated using \eqref{YuleWalker}. It is a simple exercise to check that the minimization problem in \eqref{I_bar} is convex. We are now ready to state our main result, a lower bound to the mismatch capacity associated with the system model described in Section~\ref{sec:problemFormulation}. The proof of the following result is given in Section~\ref{sec:proof}.

\begin{theorem}\label{th:1}
Consider the Gaussian ISI channel model in \eqref{2tapISI}, and the mismatched decoding metric in \eqref{VGaussian}. Then, $C_{\mathrm{ISI}}^{\mathrm{Mis}}>\max_{\bvarphit\in \calP}\bar{I}_1(\balpha,\bvarphi)$, where $\bar{I}_1(\balpha,\bvarphi)$ is given in \eqref{I_bar}.
\end{theorem}

As was mentioned in the Introduction, previous works on the mismatch capacity focused mainly on standard random coding ensembles, where each codeword is independently and identically generated according to some given probability distribution. However, since the channel has memory, it is reasonable to consider ensembles over which there is a correlation between the symbols within each codeword. To achieve $\bar{I}_1(\balpha,\bvarphi)$, for a given $\bvarphi$, the codebook $\mathfrak{C}_n$ is generated as follows: For each message $i\in[1:e^{nR}]$, we generate (independently) the sequence $\ppp{X_t}_{t=1}^n$ according to 
\begin{align}
X_t = \begin{cases}
\sum_{i=1}^p\varphi_i\cdot X_{t-i} + \eta Z_t, &t\geq1\\
0, &t<1
\end{cases}
\label{ARmodel}
\end{align}
where $\ppp{Z_t}_t$ is white noise, $\eta^2$ is chosen such that $\bE X_t^2=P_X$, for $t=[1:n]$, and thus it is given in \eqref{calculatenoisevar}. Since we assume that the roots of the polynomial $z^{p}-\sum _{i=1}^{p}\varphi_{i}z^{p-i}$ lie strictly within the unit circle, the above process is wide-sense stationary. For example, if $p=1$, we get
\begin{align}
X_t = \varphi_1\cdot X_{t-1} + \sqrt{P_X\p{1-\varphi_1^2}}Z_t,\label{ARmodel1}
\end{align}
for $t\geq1$.

The role of $\bvarphi$ is to shape the spectrum of the input process so as to mitigate the (undesired) effects of the mismatch decoding. Note, however, that any choice of $\ppp{\varphi_m}_{m=1}^p\in\calP$ would result in a legitimate lower bound to the mismatch capacity. Generally speaking, our result can be interpreted as follows: the first term in \eqref{I_bar} is associated with the differential entropy of the input process, and the second term corresponds to a certain conditional entropy of the input given the incorrectly processed output. Finally, we mention \cite[Th. 2]{abou_faycal}, where the Gaussian i.i.d. ensemble was studied using a different approach. It can be shown that, when specialized to the i.i.d. case (i.e., $\varphi_i=0$ for all $i\in[1:p]$), our result in Theorem~\ref{th:1} coincides with \cite[Th. 2]{abou_faycal}.

We now study the fixed composition ensemble, where codewords are drawn uniformly at random within the a type class of real-valued sequences. Specifically, fix an arbitrary $\varepsilon>0$, and pick $p\in\mathbb{N}$. Let $\Gamma$ denote the set of all vectors $\bgamma = (\gamma_0,\gamma_1,\ldots,\gamma_p)\in\mathbb{R}^{p+1}$ with $\gamma_0\triangleq P_X$, such that the matrix $\ppp{\gamma_{\abs{i-j}}}_{i,j}$ is a positive-definite Toeplitz matrix. Also, let $\gamma_{-k}=\gamma_k$, for $k=[1:p]$. Define the sequence of sets $\mathcal{T}^n_\varepsilon(\bgamma)$, for $n=1,2,\ldots$, as follows
\begin{align}
\mathcal{T}^n_\varepsilon(\bgamma) &\triangleq \left\{\bx\in\mathbb{R}^n:\;\abs{\frac{1}{n}\sum_{t=1}^n x_tx_{t-k}-\gamma_k}<\varepsilon,\ k=[0:p]\right\}.\label{condtypemar}
\end{align}
The codebook $\mathfrak{C}_n$ is generated by drawing $M$ codewords independently and uniformly at random from $\mathcal{T}^n_\varepsilon(\bgamma)$. The role of $\bgamma$ is to shape the spectrum of the input process, and accordingly, these parameters can be optimized. To state our main result we need some additional definitions. Let $\Pi_{m}(\balpha)\triangleq \sum_{k=0}^{K-m}\alpha_k\alpha_{k+m}$, for $m\in[1:K]$, and 
\begin{align}
\calW\triangleq\left\{\ppp{\omega_k}_{k=0}^{p+1}\in\mathbb{R}_+^{p+1}:\;\omega_0\geq\sum_{i=1}^p\omega_i+\omega_{p+1}\cdot\sum_{m=1+p\wedge K}^K\abs{\Pi_{m}(\balpha)}\right\}.
\end{align}
For $\nu \in [0,2\pi]$, and $\bomega\in\calW$, define
\begin{align}
g_{\bomegat}(\nu) \triangleq \sum_{k = 0}^p \omega_k \cos(k\nu)+\omega_{p+1}\cdot\sum_{k = 1+p\wedge K}^K\Pi_{k}(\balpha)\cos(k\nu),\label{gomegadef}
\end{align}
and
\begin{align}
\bar{I}_2(\balpha,\bgamma) &\triangleq \frac{1}{2} \log(2e \eta^2) - \min_{\bomegat \in {\calW}} \left\{\vphantom{\pp{\sum_{k=0}^K \sum_{i=0}^K \alpha_k h_i\gamma_{k-i}-\sum_{l=1+p\wedge K}^K\sum_{m=0}^{K-l}\alpha_{m}\alpha_{m+l}\gamma_l}}-\frac{1}{4\pi} \int_{0}^{2\pi} \log g_{\bomegat}(\nu) d\nu + \frac{\omega_{p+1}^2}{8\pi}\int_{\pp{0,2\pi}}\mathrm{d}\nu\frac{S_Y(\nu)\abs{A(\nu)}^2}{g_{\bomegat}(\nu)}\right. \nonumber \\
& \hspace{3.5cm} \left.-\omega_{p+1}\pp{\sum_{k=0}^K \sum_{i=0}^K \alpha_k h_i\gamma_{k-i}-\sum_{l=1+p\wedge K}^K\Pi_{l}(\balpha)\gamma_l} + \sum_{k= 0}^p \omega_k\gamma_k\right\}\label{eq:achiev_fixed_comp}
\end{align}
where $\eta^2$ above is calculated by first solving the Yule-Walker equations in \eqref{YuleWalker} to find $\bvarphi$, and then substituting in \eqref{calculatenoisevar}. Also, as before, when $p<K$, $\gamma_k$ for $k=p+1,\ldots,K$, are calculated using \eqref{YuleWalker}. It can be shown that the minimization problem in \eqref{I_bar} is convex. The following result is proved in Appendix~\ref{sec:proof2}.
\begin{theorem}\label{th:2}
Consider the Gaussian ISI channel model in \eqref{2tapISI}, and the mismatched decoding metric in \eqref{VGaussian}. Then, $C_{\mathrm{ISI}}^{\mathrm{Mis}}>\max_{\bgammat\in\Gamma}\bar{I}_2(\balpha,\bgamma)$, where $\bar{I}_2(\balpha,\bgamma)$ is given in \eqref{eq:achiev_fixed_comp}.
\end{theorem}

Comparing the achievable rates in Theorems~\ref{th:1} and \ref{th:2}, we see that the achievable rate expression in \eqref{I_bar} is expressed in terms of a one-dimensional optimization over $\omega$, while the expression in \eqref{eq:achiev_fixed_comp} is expressed in terms of a minimization over a $(p+2)$-length vector $\bomega$. Indeed, as shall be seen in the proof of Theorem~\ref{th:2} the additional parameters $\ppp{\omega_k}_{k=0}^p$ correspond to the $p+1$ constraints imposed by the codebook itself.

As mentioned in the Introduction, a byproduct of our analysis is an ensemble-tight characterization of the random coding error exponent. Indeed, the proofs of Theorems~\ref{th:1} and \ref{th:2} are based on an exponentially-tight analysis of the (ensemble) average probability of error. This implies also that the achievable rate in Theorem~\ref{th:1} is ensemble-tight, namely, one cannot achieve higher rates using the same random coding ensemble. We next characterize the random coding error exponent in the special case of $\alpha_k=0$ for $1\leq k\leq K$ (namely, a memoryless decoding metric), and we use the fixed composition ensemble in \eqref{condtypemar} with $p=0$. We emphasize that using the same methods we can analyze the more general case, but for simplicity we opted to focus on the above configuration. Let $E(P_X,R)$ designate the random coding error exponent associated with the above setting, namely, $E(P_X,R)\triangleq\liminf_{n\to\infty}-\frac{1}{n}\log\bar{P}_e(n,R)$ where $\bar{P}_e(n,R)$ designates the ensemble average probability of error. For brevity, we omit the dependency of $E(P_X,R)$ on $\alpha_0$, $\ppp{h_k}_{k=0}^{K}$, and $\sigma^2$. We start with some definitions. Let $\Pi_{m}(\bh)\triangleq \sum_{k=0}^{K-m}h_kh_{k+m}$, for $m\in[1:K]$, and for $\nu \in [0,2\pi]$, define
\begin{align}
u_{\hat{\bomegat}}(\nu) \triangleq\frac{1}{4}\abs{\hat{\omega}_1-h_0+\sum_{k=1}^Kh_ke^{-jk\nu}}^2-\p{\frac{1}{2\sigma^2}+\hat{\omega}_2}\cdot\pp{\hat{\omega}_0+\frac{1}{2}\norm{\bh}_2^2+\sum_{k=1}^K\Pi_{k}(\bh)\cos(k\nu)},\label{uomegadef}
\end{align}
where $\hat{\bomega}\in\hat{\calW}$, and $\hat{\calW}$ is the set of all $\hat{\bomega}$ satisfying $u_{\hat{\bomegat}}(\nu)>0$, for any $\nu \in [0,2\pi]$. Also, for $P_Y\in\mathbb{R}_+$, $\abs{\rho_{XY}}\leq1$, let
\begin{align}
V(\hat{\bomega},P_Y,\rho_{XY})\triangleq\frac{1}{4\pi}\int_0^{2\pi}\log \pp{4eP_X\sigma^2u_{\hat{\bomegat}}(\nu)}\mathrm{d}\nu-\hat{\omega}_0\cdot P_X-\hat{\omega}_1\cdot \rho_{XY}\sqrt{P_XP_Y}-\hat{\omega}_2\cdot P_Y.
\end{align}
Finally, for $\abs{\beta}<1$, define
\begin{align}
I(\beta,h_0,\alpha_0)\triangleq -\frac{\abs{\mathrm{sgn}(h_0)+\mathrm{sgn}(\alpha_0)}}{4}\log\p{1-\beta^2},\label{IdefExpo}
\end{align}
and note that if $h_0$ and $\alpha_0$ share the same sign then $I(\beta,h_0,\alpha_0) = -0.5\cdot\log\p{1-\beta^2}$. Otherwise, $I(\beta,h_0,\alpha_0)=0$. The following result is proved in Appendix~\ref{app:3}.
\begin{theorem}\label{th:3}
Consider the Gaussian ISI channel model in \eqref{2tapISI}, and the mismatched decoding metric in \eqref{VGaussian}, with $\alpha_k=0$, for $1\leq k\leq K$. Then,
\begin{align}
E(P_X,R) = \min_{P_Y,\rho_{XY}}\max_{\hat{\bomegat}\in\hat{\calW}}\ppp{V(\hat{\bomega},P_Y,\rho_{XY})+\pp{I(\rho_{XY},h_0,\alpha_0)-R}_+}.\label{errorExpCh}
\end{align}
\end{theorem}

Based on Theorem~\ref{th:3} we can see that if $h_0$ and $\alpha_0$ have different signs then $E(P_X,R)=0$, for any $R\geq0$ and any $P_X$. This is indeed reasonable due to the following reason: if, for example, $h_0>0$ but $\alpha_0<0$, then the mismatched decoder simply looks for the codeword which \emph{minimizes} its empirical correlation with the output sequence $\by$. However, this is exactly the opposite operation of the optimal ML decoder which maximizes the empirical correlation with $\by$. Also, one can argue that \ref{errorExpCh} resembles the famous Csisz\'ar-K\"{o}rner-style error exponent function \cite{CsisKro}, namely, $\min_Q\{D_{\mathrm{KL}}(Q||P)+\pp{I(Q)-R}_+\}$, where $D_{\mathrm{KL}}(Q||P)$ is the Kullback-Leibler (KL) divergence between two measures $Q$ and $P$, and $I(Q)$ is the mutual information calculated w.r.t. $Q$. This indeed makes sense, and one can think of $\max_{\hat{\bomegat}\in\hat{\calW}}V(\hat{\bomega},P_Y,\rho_{XY})$ in \eqref{errorExpCh} as playing the role analogous to the KL-divergence. For example, if $K=0$, then it is a simple task to check that
\begin{align}
\max_{\hat{\bomegat}\in\hat{\calW}}V(\hat{\bomega},P_Y,\rho_{XY}) = -\frac{1}{2}\log\frac{(1-\rho_{XY}^2)P_Y}{\sigma^2}+\frac{1}{2\sigma^2}\pp{P_Y-2h_0\rho_{XY}\sqrt{P_XP_Y}+h_0^2P_X}-\frac{1}{2}
\end{align}
which is just the KL-divergence $D_{\mathrm{KL}}(Q_{XY}||P_{XY})$, with $Q_{XY}$ and $P_{XY}$ both being multivariate Gaussian distributions, with zero means, and covariances
\begin{align}
\mathbf{\Sigma}_Q = \begin{bmatrix}
    P_X & \rho_{XY}\sqrt{P_XP_Y}\\
		\rho_{XY}\sqrt{P_XP_Y} & P_Y
\end{bmatrix}\ \ \ \mathrm{and}\ \ \
\mathbf{\Sigma}_P = \begin{bmatrix}
    P_X & h_0\cdot P_X\\
		h_0\cdot P_X & h_0^2\cdot P_X+\sigma^2
\end{bmatrix},
\end{align}
respectively. In general, the term $\max_{\hat{\bomegat}\in\hat{\calW}}V(\hat{\bomega},P_Y,\rho_{XY})$ can be thought as the asymptotic formula of the $n$-letter weighted KL-divergence $D_{\mathrm{KL}}(Q_{Y^n|X^n}||W_{Y^n|X^n}\vert\mu_{X^n})$, where $Q_{Y^n|X^n}$ is some test channel, and $\mu_{X^n}$ is the random coding distribution, i.e., a uniform measure over the $n$-dimensional hypersphere with radius $\sqrt{nP_X}$.

\begin{figure}[!t]
\begin{minipage}[b]{1.0\linewidth}
\centering
\centerline{\includegraphics[width=14cm,height = 10cm]{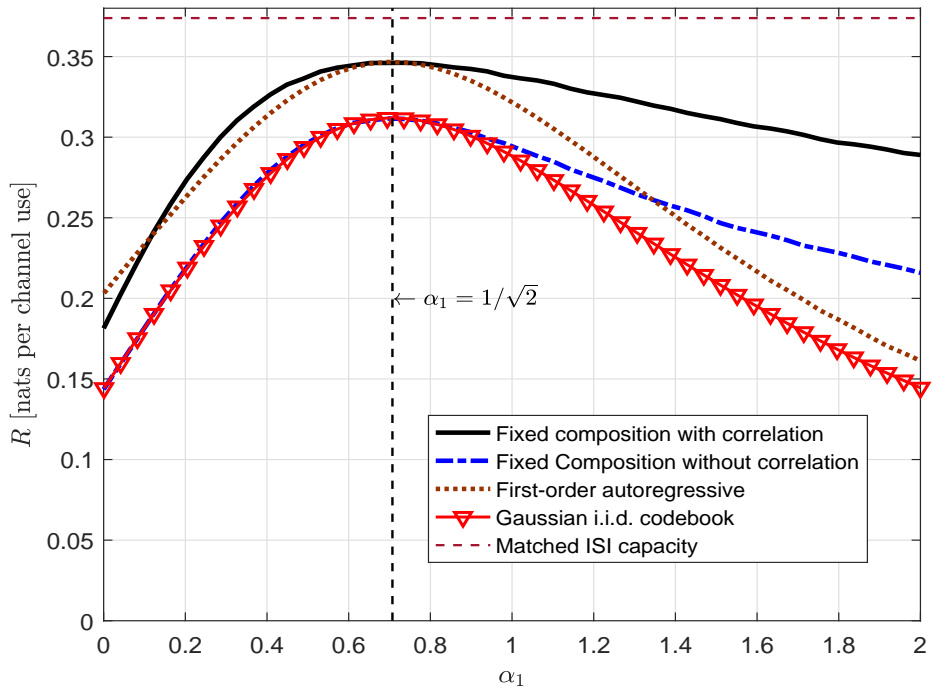}}
\end{minipage}
\caption{Achievable rates as a function of the mismatched level $\alpha_1$, for the Gaussian ISI channel with $h_0=h_1=1/\sqrt{2}$, $\sigma^2=1$, $P_X=1$, and $\alpha_0=1/\sqrt{2}$. The matched capacity is $C=0.374$ nats per channel use. The achievable rates correspond to fixed composition ensemble with one correlation parameter (solid black curve), fixed composition ensemble without correlations (dashed-dotted blue curve), first-order autoregressive ensemble (dotted brown curve), and Gaussian i.i.d. codebook (solid $v$-mark red curve).}
\label{fig:2}
\end{figure}

We next compare numerically the results obtained in Theorems~\ref{th:1} and \ref{th:2}. Fig.~\ref{fig:2} presents a numerical comparison of the results obtained in Theorems~\ref{th:1} and \ref{th:2}, in the following scenario. We consider the two-tap Gaussian ISI channel with $h_0=h_1=1/\sqrt{2}$, $\sigma^2=1$, and $P_X=1$. The mismatched decoder has a fixed coefficient $\alpha_0=1/\sqrt{2}$, and we calculate the achievable rates as a function of $\alpha_1$. The matched capacity can be calculated numerically and it is given by $C=0.374$ nats per channel use. The achievable rates in Fig.~\ref{fig:2} correspond to fixed composition ensemble with one correlation parameter (solid black curve), fixed composition ensemble without correlations (dashed-dotted blue curve), first-order autoregressive ensemble (dashed brown curve), and Gaussian i.i.d. codebook (solid $v$-marked red curve). It can be seen that the fixed composition ensemble with one correlation parameter is almost uniformly better than all the other ensembles. The Gaussian i.i.d. codebook is the worst ensemble. In this example, all ensembles achieve the maximum rate at the matched value of $\alpha_1$. Interestingly, we see that all ensembles (and especially the fixed composition ensemble with correlation) behave differently in the regions $\alpha_1<1/\sqrt{2}$ and $\alpha_1>1/\sqrt{2}$. These regions, respectively, correspond to the ``optimistic" and ``pessimistic" assumptions of the receiver regarding the ISI channel. To wit, when $\alpha_1<1/\sqrt{2}$ ($\alpha_1>1/\sqrt{2}$) the receiver can be thought of as being optimistic (pessimistic) since he assumes that the ISI part is weaker (stronger) than what it really is. Accordingly, it seems that in terms of achievable rates, the price of optimism is higher than the price of pessimism. 

\begin{figure}[!t]
\begin{minipage}[b]{1.0\linewidth}
\centering
\centerline{\includegraphics[width=14cm,height = 10cm]{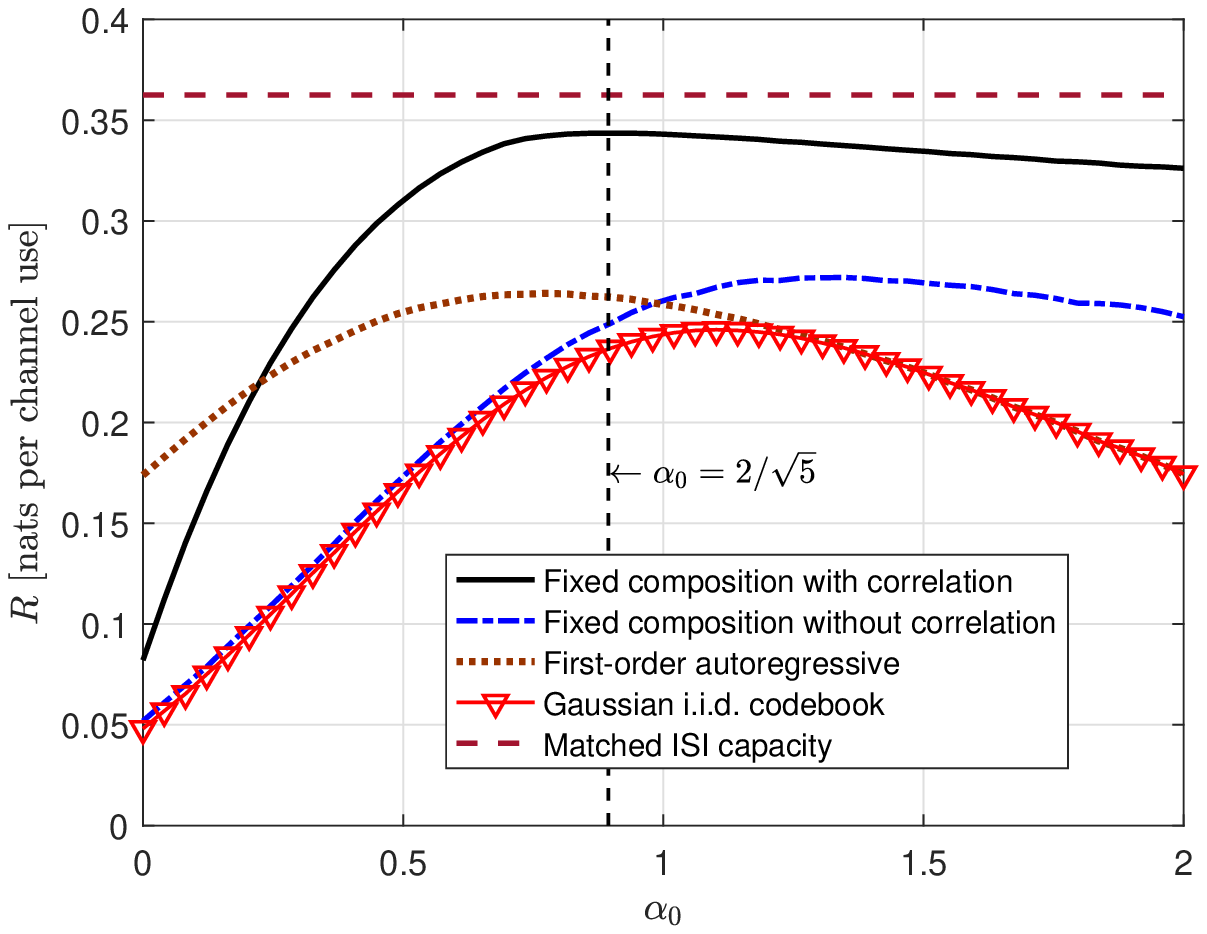}}
\end{minipage}
\caption{Achievable rates as a function of the mismatched level $\alpha_0$, for the Gaussian ISI channel with $h_0=2/\sqrt{5}$, $h_1=1/\sqrt{5}$, $\sigma^2=1$, $P_X=1$, and $\alpha_1=1$. The matched capacity is $C=0.3625$ nats per channel use. The achievable rates correspond to fixed composition ensemble with one correlation parameter (solid black curve), fixed composition ensemble without correlations (dashed-dotted blue curve), first-order autoregressive ensemble (dotted brown curve), and Gaussian i.i.d. codebook (solid $v$-mark red curve).}
\label{fig:3}
\end{figure}

\begin{figure}[!t]
\begin{minipage}[b]{1.0\linewidth}
\centering
\centerline{\includegraphics[width=14cm,height = 9cm]{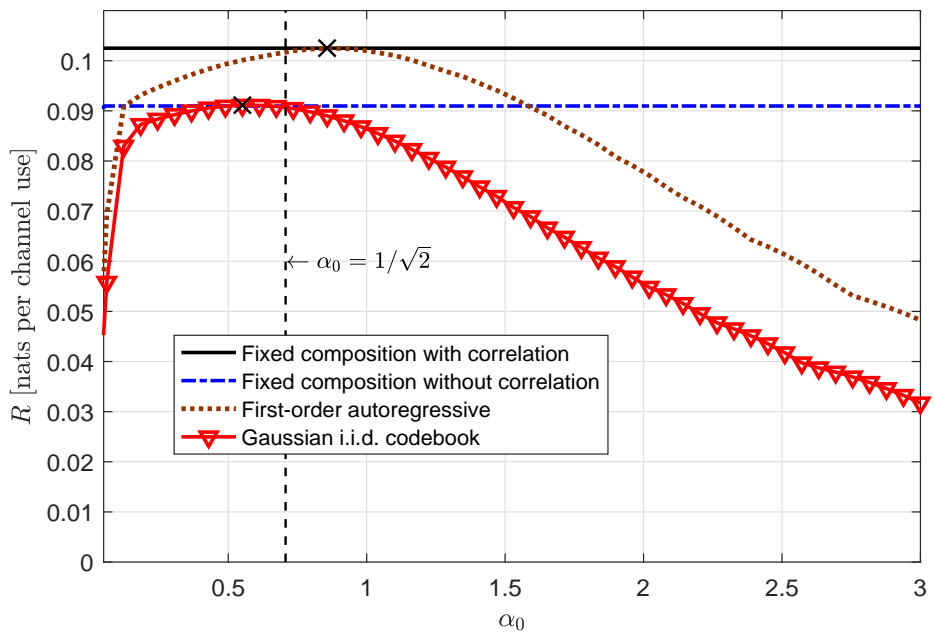}}
\end{minipage}
\caption{Achievable rates as a function of the mismatched level $\alpha_0$, for the Gaussian ISI channel with $h_0=h_1=h_2=1/\sqrt{2}$, $\sigma^2=1$, $P_X=1$. The achievable rates correspond to fixed composition ensemble with one correlation parameter (solid black curve), fixed composition ensemble without correlations (dashed-dotted blue curve), first-order autoregressive ensemble (dotted brown curve), and Gaussian i.i.d. codebook (solid $v$-mark red curve). The cross symbols ``X" refer to the maximum rate achieved by the corresponding ensembles.}
\label{fig:5}
\end{figure}

Another comparison is shown in Fig.~\ref{fig:3}, where now $h_0=2/\sqrt{5}$, $h_1=1/\sqrt{5}$, $\sigma^2=1$, $P_X=1$, and $\alpha_1=1$. The matched ISI capacity in this case is $C=0.3625$ nats per channel use. In this case, since $\alpha_1\neq h_1$, the mismatched channel is never the same as the true channel. Contrary to Fig.~\ref{fig:2}, it can be seen that here the maximum rate is not achieved at the matched value of $\alpha_0$. This observation illustrates that, in general, there is no direct connection between achievable rates and the minimization of the $\ell_2$ distance between the true channel and the mismatched channel. Instead, if one imposes a fixed $\ell_2$ distance between the true response and the mismatched response, the achievable rates might very well depend of how well the mismatched response approximate the ``good" part of the frequency domain representation of the true response. Similarly, in Fig.~\ref{fig:5} we consider the case where $h_0=h_1=h_2=1/\sqrt{2}$, $\sigma^2=1$, and $P_X=1$. The mismatched decoder has only one coefficient $\alpha_0$ (namely, $K=0$). Here, again, it can be seen that the maximum rate is not achieved at $\alpha_0 = 1/\sqrt{2}$. This illustrates that truncation (of the mismatched decoder) is not always optimal. Note that this result might be initially surprising since one can argue that, at least in the case of Gaussian i.i.d. codebook, if one decides to ignore the contribution of the ISI part, since input symbols are independent of each other, then this contribution plays the role of additional Gaussian additive noise. Accordingly, the best choice of the mismatch parameter should be the truncated one (namely $\alpha_0=h_0$). This intuition is misleading due to the same reason mentioned above.

\subsection{Mismatched Universal Decoders}\label{subsec:misdec}

As discussed in the Introduction in many situations in coded communication systems, channel uncertainty and variability preclude the implementation of the optimum ML decoder. In such cases, a good solution is provided by universal decoders
which perform asymptotically as well as the ML decoder and yet do not require knowledge of the channel. 

In this subsection, we analyze the following scenario: consider the same channel model presented in Section~\ref{sec:problemFormulation}. Then, assume that due to complexity issues concerning the implementation of the optimal ML decoder, the receiver uses the mismatched decoding metric in \eqref{VGaussian} with only $\alpha_0$ being active, i.e., corresponding to a memoryless channel. Nonetheless, we allow our receiver to optimize this coefficient, namely, it can be a function of the true channel. Now, consider a different receiver which uses a universal decoder designed for a memoryless channel (namely without ISI). In other words, the true family of channels is \emph{outside} the class of channels for which the universal decoder is actually designed, i.e, mismatched universal decoder. Then, which approach yields higher rates?

In the sequel, for simplicity, we focus on the fixed composition ensemble without correlations. It is well-known (see, e.g., \cite{MerhavLapidoth,Wasim3,FerderLapidoth,UniNeri2,MerhavArbitrary}) that, for the AWGN with codewords drawn uniformly and independently over the $n$-dimensional hypersphere with radius $\sqrt{nP_X}$, given an output sequence $y^n$, the decoder
\begin{align}
\hat{i}=\arg\max_{i\in\mathfrak{C}_n}\abs{\sum_{t=1}^nx_{t,i}y_t},\label{GLRT}
\end{align}
is universal. Indeed, in this scenario, the generalized likelihood-ratio test (GLRT) is universal, and it is a simple exercise to check that the GLRT is equivalent to \eqref{GLRT}. Before we present our main result, we briefly comment that using the same techniques developed in this paper, one can consider more complicated ensembles, such as the one in \eqref{condtypemar}, and universal decoders designed for the ISI channel \cite{Wasim3} and not just for the AWGN, as described above. In principle, it makes sense that the random coding ensemble and the universal decoder would be consistent with one another in the sense of their assumptions on the memorylessness/memoryfulness of the channel. However, in the following discussion we demonstrate that we can consider the ensemble in \eqref{condtypemar} when the universal decoder is designed for a memoryless channel, with the understanding that we can actually consider also universal decoders designed for the ISI channel \cite{Wasim3}. Given $\varepsilon>0$, and two sequences $x^n$ and $y^n$, we define
\begin{align}
\calT_\varepsilon^n(\bx\vert\by)\triangleq\ppp{\bx'\in\mathbb{R}^n:\;\abs{\frac{1}{n}\sum_{t=1}^nx_t'y_t-\frac{1}{n}\sum_{t=1}^nx_ty_t}<\epsilon}.
\end{align}
Let $\mu(\cdot)$ designate the uniform measure over \eqref{condtypemar}, and define
\begin{align}
U(\bx,\by) &\triangleq -\frac{1}{n}\log \mu\p{\calT_\varepsilon^n(\bx\vert\by)}\\
& = \frac{1}{n}\log\mathrm{Vol}\p{\mathcal{T}^n_\varepsilon(\bgamma)}-\frac{1}{n}\log\mathrm{Vol}\p{\mathcal{T}^n_\varepsilon(\bgamma)\cap\calT_\varepsilon^n(\bx\vert\by)}.\label{unimetric}
\end{align}
Then, it can be shown that the decoder which maximizes the metric in \eqref{unimetric} over the given codebook, is universal w.r.t. the ensemble in \eqref{condtypemar}. Since the first term at the right hand side (r.h.s.) of \eqref{unimetric} is in fact independent of $\bx$ it can be omitted, and so to explicitly derive the above decoding metric, we just need to evaluate the log-volume of the set $\mathcal{T}^n_\varepsilon(\bgamma)\cap\calT_\varepsilon^n(\bx\vert\by)$. This can be done using the same techniques used in the proof of Theorem~\ref{th:2}. Returning back to our setting, we have the following result.
\begin{theorem}\label{th:4}
Consider the Gaussian ISI channel model in \eqref{2tapISI}, and the mismatched universal decoder in \eqref{GLRT}. Then, every
\begin{align}
R<\frac{1}{2}\log\p{1+\frac{h_0^2\cdot P_X}{(\norm{\bh}_2^2-h_0^2)\cdot P_X+\sigma^2}}\label{achuniversal}
\end{align}
is achievable.
\end{theorem}
As before, we can obtain the random coding error exponent associated with the above universal decoder.
\begin{theorem}\label{cor:1}
Consider the Gaussian ISI channel model in \eqref{2tapISI}, and the mismatched universal decoder in \eqref{GLRT}. Then,
\begin{align}
E(P_X,R) = \min_{P_Y,\rho_{XY}}\max_{\hat{\bomegat}\in\hat{\calW}}\ppp{V(\hat{\bomega},P_Y,\rho_{XY})+\pp{-\frac{1}{2}\log(1-\rho_{XY}^2)-R}_+}.
\end{align}
\end{theorem}

\begin{figure}[!t]
\begin{minipage}[b]{1.0\linewidth}
\centering
\centerline{\includegraphics[width=11cm,height = 8cm]{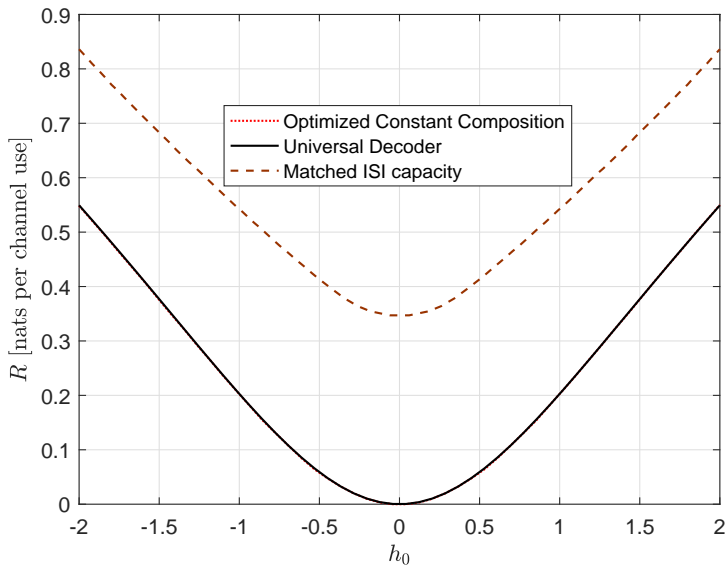}}
\end{minipage}
\caption{Achievable rates as a function of the coefficient $h_0$, for the Gaussian ISI channel with two taps and $h_1=1$, $\sigma^2=1$, and $P_X=1$. The achievable rates correspond to the optimized (over $\alpha_0$) mismatched decoder (dotted red curve), and the mismatched universal decoder (solid black curve), both under fixed composition ensemble without correlations.}
\label{fig:4}
\end{figure}

We provide a proof sketch of Theorem~\ref{th:4} in Appendix~\ref{app:4}. The proof of Theorem~\ref{cor:1} is essentially the same as the proof of Theorem~\ref{th:3}, and thus omitted. From Theorem~\ref{th:4} we see that the achievable rate in \eqref{achuniversal} has the interpretation that the ``mismatched input signal" (or the residue signal) is treated as additional Gaussian noise at the decoder. Accordingly, it is interesting to compare the above results with Theorems~\ref{th:2} and \ref{th:3}. Some technical calculations reveal that, for the specific scenario we consider above, the achievable rate $\max_{\alpha_0}\bar{I}_2(\alpha_0,P_X)$ in Theorem~\ref{th:2} is essentially \emph{exactly} the same as \eqref{achuniversal}. Moreover, the optimal $\alpha_0$ which achieves \eqref{achuniversal} should be chosen as to match the sign of $h_0$, namely, if $h_0>0$ then $\alpha_0$ can be any real but positive number, while if $h_0<0$ then $\alpha_0$ can be any real but negative number. Accordingly, while the mismatched decoder needs some knowledge of the true channel, it achieves the same rates as the universal decoder which has no knowledge at all about the true channel. In particular, if the sign of $\alpha_0$ was unknown to the mismatched decoder, and the mismatch decoder mistakenly assumes a different sign compared to the true channel, then it will achieve zero-rate. This basically demonstrates that, at least in the specific scenario described above, our results provide indications that universal decoders exhibit a robustness property w.r.t. the family of channels over which they are actually designed. To wit, even though our universal decoder is designed for a completely different class of channels, it still performs as well as the best mismatched ML decoder in the same family. This observation (potentially) suggests a certain expansion of the classic notion of universality to cases where the true underlying channel is outside the class. Similarly, comparing the error exponents in Theorem~\ref{th:3} and \ref{cor:1} we see that the only difference is the additional $\abs{\mathrm{sgn}(h_0)+\mathrm{sgn}(\alpha_0)}/2$ term in \eqref{IdefExpo}. Accordingly, if $h_0$ and $\alpha_0$ share the same sign, then both exponents coincide, otherwise, the exponent in Theorem~\ref{th:3} vanishes, while the exponent in Theorem~\ref{cor:1} remains unaffected. Finally, we present in Fig.~\ref{fig:4} a numerical calculation of the achievable rate in Theorem~\ref{th:4}, for the two-tap Gaussian ISI channel with $h_1=1$, $\sigma^2=1$, and $P_X=1$. 

\section{Proof of Theorem~\ref{th:1}}\label{sec:proof}

The codebook $\mathfrak{C}_n=\ppp{\bx_i}_{i=1}^M$ is generated as follows: For each message $i\in[1:e^{nR}]$, we generate (independently) a sequence $\ppp{X_{t}}_{t=1}^n$ according to \eqref{ARmodel}, where $\ppp{Z_t}_t$ is white Gaussian noise, $\eta^2$ is chosen such that $\bE X_t^2=P_X$, for any $t=[1:n]$, and thus it is given in \eqref{calculatenoisevar}. Also, for technical reasons, we fix $X_t = 0$ for $t=n+1,\ldots,n+(p\vee K)$, and for each message $i\in[1:e^{nR}]$. We will then think of all the forthcoming summation terms over $t=[1:n]$ as summing up to $\bar{n}=n+p\vee K$ instead of $n$. This assumption is made only for convenience, but has no influence on either the achievable error exponents or the achievable rates, as long as $p\vee K$ is fixed and $n\to\infty$. Accordingly, we can write the following circularity relation
\begin{align}
\sum_{t=1}^{\bar{n}}X_{t-i}X_{t-j} = \sum_{t=1}^{\bar{n}}X_{t}X_{t-(j-i)}\label{circularAssump}
\end{align}
for $i\leq j\in\pp{1:p\vee K}$. For notational convenience, in the following we use $n$ in place of $\bar{n}$. We let $\mu_p(\cdot)$ designate the probability density function of a sequence of RV's generated as described above. Finally, define
\begin{align}
\gamma_m\triangleq\bE\ppp{X_tX_{t-m}}\label{autocovariance}
\end{align}
to be the auto-covariance of the autoregressive process, and recall that these coefficients can be found by evaluating \eqref{YuleWalker}.

Without loss of generality, we assume throughout, that the transmitted codeword is $\bx\triangleq\bx_1$. Accordingly, the average probability of error can be written as 
\begin{align}
\bar{P}_{e}(n,R) &= \Pr\pp{\bigcup_{i=2}^{M}\ppp{\frac{1}{n}\log V\p{\bY\vert \bX_{i}}\geq \frac{1}{n}\log V\p{\bY\vert \bX_1}}}\label{OrignalProb}\\
& = \bE\ppp{\left.\Pr\pp{\bigcup_{i=2}^{M}\ppp{\frac{1}{n}\log V\p{\bY\vert \bX_{i}}\geq \frac{1}{n}\log V\p{\bY\vert \bX_1}}\right\vert\calF_0}}\label{innerPro}
\end{align}
where $\calF_0 \triangleq \p{\bX_1,\bY}$. Recall that for a pairwise independent events $\ppp{\calA_i}_{i=1}^M$, we have \cite[Sec. A2]{Shulman}
$$
\frac{1}{2}\min\ppp{1,\sum_{i=1}^M\Pr\ppp{\calA_i}}\leq\Pr\ppp{\bigcup_{i=1}^M\calA_i}\leq \min\ppp{1,\sum_{i=1}^M\Pr\ppp{\calA_i}},
$$
and therefore,
\begin{align}
\bar{P}_{e}(n,R) \doteq\bE\pp{\min\ppp{1,M\cdot\Pr\ppp{\left.\frac{1}{n}\log V\p{\bY\vert \bX_2}\geq \frac{1}{n}\log V\p{\bY\vert \bX_1}\right\vert\calF_0}}}. \label{ShulmaninnerPro}
\end{align}
Thus, we need to assess the exponential behavior of the probability term in \eqref{ShulmaninnerPro}. Fix a positive constant $B>0$, and define the sequence of sets
\begin{align}
\calH_n(B) \triangleq \ppp{\bx,\by\in\mathbb{R}^n:\;\frac{1}{n}\sum_{i=1}^nx_i^2\leq B,\;\frac{1}{n}\sum_{i=1}^ny_i^2\leq B},\label{Hdef0}
\end{align}
for $n\geq1$. Accordingly, \eqref{ShulmaninnerPro} can be rewritten as follows
\begin{align}
\bar{P}_{e}(n,R) &\doteq\bE\pp{\calI\ppp{\calH_n(B)}\cdot\min\ppp{1,M\cdot\Pr\ppp{\left.\frac{1}{n}\log V\p{\bY\vert \bX_2}\geq \frac{1}{n}\log V\p{\bY\vert \bX_1}\right\vert\calF_0}}}\nonumber\\
&+\bE\pp{\calI\ppp{\calH^c_n(B)}\cdot\min\ppp{1,M\cdot\Pr\ppp{\left.\frac{1}{n}\log V\p{\bY\vert \bX_2}\geq \frac{1}{n}\log V\p{\bY\vert \bX_1}\right\vert\calF_0}}}. \label{ShulmaninnerPro20}
\end{align}
Using the same techniques as in \cite[Lemma 2]{NeriUni} and \cite[Lemma 2]{Wasim3} it can be shown that there exists a sufficiently large $B$ such that the first term at the r.h.s. of dominates in the exponential scale. This passage is mainly technical, and will be used to evaluate the volume of some typical sets as shall be clear in the sequel. To conclude, we have
\begin{align}
\bar{P}_{e}(n,R) &\doteq\bE\pp{\calI\ppp{\calH_n(B)}\cdot\min\ppp{1,M\cdot\Pr\ppp{\left.\frac{1}{n}\log V\p{\bY\vert \bX_2}\geq \frac{1}{n}\log V\p{\bY\vert \bX_1}\right\vert\calF_0}}}.\label{barPenRSurvive}
\end{align}
For $(\bX_1,\bY) = (\bx,\by)\in\calH_n(B)$, the inner probability term at the r.h.s. of \eqref{barPenRSurvive} can be represented as
\begin{align}
\Pr\ppp{\frac{1}{n}\log V\p{\by\vert \bX_2}\geq \frac{1}{n}\log V\p{\by\vert \bx}} = \int_{\bxt'\in\calV(\bxt,\byt)} \mathrm{d}\bx'\mu_p(\bx')\label{lowUppermar}
\end{align}
where
\begin{align}
\calV(\bxt,\byt)\triangleq\ppp{\bx'\in\mathbb{R}^n:\;\frac{1}{n}\log V\p{\by\vert \bx'}\geq \frac{1}{n}\log V\p{\by\vert \bx}}.
\end{align}
 
Using the saddle-point integration method we asses the exponential behavior of the r.h.s. of \eqref{lowUppermar}. To present the result, let $\omega\in\mathbb{R}_+$, and define for $1\leq m\leq K$ and $1\leq r\leq p$,
\begin{align}
\Pi_{m}(\balpha)&\triangleq \sum_{k=0}^{K-m}\alpha_k\alpha_{k+m},\label{prodalphadef} \\
\hat{\Pi}_{r}(\bvarphi)&\triangleq \sum_{k=1}^{p-r}\varphi_k\varphi_{k+r},\label{prodvarphidef}
\end{align}
and
\begin{align}
\psi(\bx,\by)\triangleq \sum_{l=0}^K\alpha_l\sum_{t=1}^ny_tx_{t-l}-\frac{1}{2}\norm{\balpha}_2^2\sum_{t=1}^nx_t^{2}-\sum_{l=1}^K\Pi_l(\balpha)\sum_{t=1}^nx_{t}x_{t-l}.\label{bpsivec}
\end{align}
Also, let $\mathbf{\Omega} \triangleq \mathbf{\Omega}_1+\mathbf{\Omega}_2$ be the $n\times n$ symmetric Toeplitz matrix,
\begin{align}
\pp{\mathbf{\Omega}_1}_{i,j} \triangleq \frac{\omega}{2}\cdot\begin{cases}
\norm{\balpha}_2^2,&\ \text{for}\;\abs{i-j}=0\\
\Pi_{\abs{i-j}}(\balpha),&\ \text{for}\;1\leq \abs{i-j}\leq K\\
0, &\ \text{otherwise}
\end{cases},\label{omega1}
\end{align}
and
\begin{align}
\pp{\mathbf{\Omega}_2}_{i,j} \triangleq \frac{1}{2\eta^2}\cdot\begin{cases}
1+\norm{\bvarphi}_2^2,&\ \text{for}\;\abs{i-j}=0\\
\hat{\Pi}_{\abs{i-j}}(\bvarphi)-\varphi_{\abs{i-j}},&\ \text{for}\;1\leq\abs{i-j}\leq p\\
0, &\ \text{otherwise}
\end{cases}.\label{omega2}
\end{align} 
Let $\mathbf{A}$ be an $n\times n$ lower-triangular Toeplitz matrix where for $i\geq j$,
\begin{align}
\pp{\mathbf{A}}_{i,j} \triangleq\begin{cases}
\alpha_{i-j},&\ \text{for}\;0\leq i-j\leq K\\
0, &\ \text{otherwise}
\end{cases}.\label{Amat}
\end{align}
Finally, define
\begin{align}
q_n(\omega,\bx,\by)\triangleq \frac{1}{2}\log\pi-\frac{1}{2n}\log\det\mathbf{\Omega}+\frac{\omega^2}{4n}{\by}^T\mathbf{A}\mathbf{\Omega}^{-1}\mathbf{A}^T{\by}-\frac{\omega}{n}\psi(\bx,\by),\label{hdef}
\end{align}
and
\begin{align}
h_n(\bx,\by)\triangleq \min_{\omega\in\mathbb{R}_+}q_n(\omega,\bx,\by)-\frac{1}{2}\log(2\pi\eta^2).\label{hndef}
\end{align}
The following lemma is proved in Appendix~\ref{app:1}.
\begin{lemma}\label{lem:volume}
Fix $(\bx,\by)\in\calH_n(B)$. Then, for any $\delta_1>0$, there exists $N_{\delta_1}\in\mathbb{N}$ large enough, such that for any $n>N_{\delta_1}$,
\begin{align}
\abs{\frac{1}{n}\log\int_{\calV(\bxt,\byt)}\mathrm{d}\bx'\mu_p(\bx')-h_n(\bx,\by)}<\delta_1.\label{VolEstimate}
\end{align}
\end{lemma}

Using \eqref{barPenRSurvive}-\eqref{lowUppermar}, and Lemma~\ref{lem:volume}, we may conclude that
\begin{align}
\bar{P}_{e}(n,R) \doteq \bE\pp{\calI\ppp{\calH_n(B)}\exp\ppp{-n\pp{-h_n(\bX,\bY)-\delta_1-R}_+}}.
\label{ShulmaninnerPro3mar}
\end{align}
The next step is taking the expectation w.r.t. $(\bX,\bY)$, distributed according to $\mu_p \times W$. In the following, we calculate the limit of \eqref{hndef} as $n\to\infty$. Due to the fact that $q_n(\omega,\bx,\by)$ converges uniformly to a limit almost surely, as shall be shown in a moment, we can interchange the order of the limit and the minimization in \eqref{hndef} \cite{Folland}. Thus, we focus on the limit of $q_n(\omega,\bx,\by)$, and we consider the asymptotics of each term in \eqref{hdef}. Since $\mathbf{\Omega}$ is a Toepliz matrix, using Szeg\"{o} theorem \cite{Szego}, we have
\begin{align}
\lim_{n\to\infty}\frac{1}{n}\log\det\mathbf{\Omega} = \frac{1}{2\pi}\int_{\pp{0,2\pi}}\mathrm{d}\nu\log f_{\omega}(\nu)
\end{align}
where
\begin{align}
f_{\omega}(\nu) &\triangleq \frac{\omega}{2}\pp{\norm{\balpha}_2^2+2\sum_{k=1}^K\Pi_k(\balpha)\cos(k\nu)}+\frac{1}{2\eta^2}\pp{1+\norm{\bvarphi}_2^2+2\sum_{k=1}^{p}(\hat{\Pi}_k(\bvarphi)-\varphi_k)\cos(k\nu)}\\
& = \frac{\omega}{2}|A(\nu)|^2   +\frac{1}{2\eta^2}\left[1+ |\Phi(\nu)|^2 - 2\cdot\mathrm{Re}(\Phi(\nu))\right].
\end{align}
Next, using the law of large numbers (LLN) and once again Szeg\"{o} theorem, we get, with overwhelming probability as $n\to\infty$,
\begin{align}
\lim_{n\to\infty}\frac{1}{n}\by^T\mathbf{A}\mathbf{\Omega}^{-1}\mathbf{A}^T\by &= \lim_{n\to\infty}\frac{1}{n}\bE\p{\bY^T\mathbf{A}\mathbf{\Omega}^{-1}\mathbf{A}^T\bY}\label{LLNa}\\
& = \lim_{n\to\infty}\frac{1}{n}\tr\ppp{\mathbf{A}\mathbf{\Omega}^{-1}\mathbf{A}^T\mathbf{R}_{\byt}}\\
& = \frac{1}{2\pi}\int_{\pp{0,2\pi}}\mathrm{d}\nu\frac{S_Y(\nu)\abs{A(\nu)}^2}{f_{\omega}(\nu)}
\end{align}
where $S_X(\cdot)$, $S_Y(\cdot)$, and $f_{\omega}(\cdot)$, are defined in \eqref{eq:S_X}-\eqref{fomega}, respectively, and $S_Y(\cdot)$ is the spectral density function of $\bY$, namely, it is the Fourier transform of the auto-covariance  
\begin{align}
\mathrm{R}_Y(t,s) &= \sum_{m,k=0}^Kh_{m}h_{k}\bE\ppp{X_{t-m}X_{s-k}} + \mathrm{R}_W(t-s)\\
& = \sum_{m,k=0}^Kh_{m_1}h_{m_2}\gamma_{t-s+k-m} + \mathrm{R}_W(t-s)\\
& = \sum_{m,k=0}^Kh_{m_1}h_{m_2}\gamma_{u+k-m} + \mathrm{R}_W(u)\\
& = \mathrm{R}_Y(u),
\end{align}
where $\ppp{\gamma_m}$ are defined in \eqref{autocovariance}, and thus
\begin{align}
S_Y(\nu) &= \sum_{m,k=0}^Kh_{m_1}h_{m_2}S_X(\nu)e^{j(m-k)\nu}+\sigma^2\\
& = \pp{\sum_{m=0}^Kh_m^2+2\cdot\sum_{m<k}^Kh_mh_k\cos\pp{(k-m)\nu}}S_X(\nu)+\sigma^2.
\end{align}
Finally, by the LLN and by using \cite[Th. 2.4.2]{TimeSeriesBook} we have, with overwhelming probability as $n\to\infty$,
\begin{align}
\lim_{n\to\infty}\frac{1}{n}\psi(\bx,\by) = \sum_{l=0}^K\sum_{i=0}^K\alpha_lh_i\gamma_{l-i}-\p{\frac{1}{2}\norm{\balpha}_2^2\gamma_0+\sum_{l=1}^K\Pi_l(\balpha)\gamma_l}.\label{LLNa2}
\end{align}
Collecting the last results, we obtain that with probability approaching one as $n\to\infty$,
\begin{align}
\lim_{n\to\infty}h_n(\bX,\bY) &= \min_\omega\left\{ \frac{1}{2}\log\pi-\frac{1}{4\pi}\int_0^{2\pi}\mathrm{d}\nu\log f_\omega(\nu)+\frac{\omega^2}{8\pi}\int_{\pp{0,2\pi}}\mathrm{d}\nu\frac{S_Y(\nu)\abs{A(\nu)}^2}{f_{\omega}(\nu)}\right.\nonumber\\
&\left.\ \hspace{3em} -\omega\pp{\sum_{l=0}^K\sum_{i=0}^K\alpha_lh_i\gamma_{l-i}-\p{\frac{1}{2}\norm{\balpha}_2^2\gamma_0+\sum_{l=1}^K\Pi_l(\balpha)\gamma_l}}\right\}-\frac{1}{2}\log(2\pi\eta^2)\nonumber\\
&\triangleq -\bar{I}_1(\balpha,\bvarphi).\label{convI}
\end{align}
Next, for any $\delta_2>0$, define the set
\begin{align}
\calA_n(\delta_2)&\triangleq\ppp{(\bx,\by):\;\abs{h_n(\bx,\by)+\bar{I}_1(\balpha,\bvarphi)}<\delta_2}\label{Adef}.
\end{align}
Accordingly, from \eqref{ShulmaninnerPro3mar} we have
\begin{align}
\bar{P}_{e}(n,R) &\leq \bE\pp{\calI\ppp{\calA_n(\delta_2)}\exp\ppp{-n\pp{-h_n(\bX,\bY)-\delta_1-R}_+}}+\Pr\ppp{\calA_n^c(\delta_2)},
\label{ShulmaninnerPro3mar2}
\end{align}
and in light of \eqref{convI}, taking $\delta_1\to0$ and $\delta_2\to0$, followed by $n\to\infty$, the last term at the r.h.s. of \eqref{ShulmaninnerPro3mar2} is asymptotically negligible. Hence, in terms of achievable rate, we get that $\bar{P}_{e}(n,R)$ decays to zero as long as
\begin{align}
R<\bar{I}_1(\balpha,\bvarphi).
\end{align}
Since $\ppp{\varphi_i}_{i=1}^p\in\calP$ were fixed parameters up to this point, we take the maximum of the r.h.s. of the above inequality over these parameters, which concludes the proof.

\section{Conclusions}\label{sec:conc}
In this paper, we considered the problem of channel coding over Gaussian ISI channels with a mismatched decoding rule. 
For this problem we provided two achievable rates using an autoregressive random coding ensemble, and fixed composition ensemble. We then presented a mismatched universal decoder, designed outside the true class of channels, and showed that it is robust. Finally, we discuss a certain generalization and some implications of our analysis. Using the same techniques developed in this paper, we can analyze more complicated scenarios of multi-user systems, such as, the ISI multiple-access channel \cite{ChengVerdu}, and the ISI broadcast channel. Specifically, to obtain ensemble-tight results, the only step in our proof that should be modified is the application of the truncated union-bound, which is not necessarily tight in the multi-user settings. To this end, one can use the tighter union bounds that were derived in \cite{scarletNew,HuleihelMerhav}.

\appendices
\numberwithin{equation}{section}
\section{Proof of Lemma~\ref{lem:volume}}\label{app:1}
Recall the fact that the step function $\Ind\ppp{x\geq0}$ is the inverse Laplace transform of the function $1/s$, i.e.,
\begin{align}
\Ind\ppp{x\geq0} = \frac{1}{2\pi j}\int_{c-j\infty}^{c+j\infty}\mathrm{d}t\frac{\exp\p{tx}}{t}, \label{eq:laplace}
\end{align}
for any $c>0$. Also, note that
\begin{align}
\calV(\bx,\by) &= \ppp{\bx'\in\mathbb{R}^n:\;\log V(\by|\bx')\geq\log V(\by|\bx)}\nonumber	\\
& = \ppp{\bx'\in\mathbb{R}^n:\;\sum_{l=0}^K\alpha_l\sum_{t=1}^ny_t(x_{t-l}'-x_{t-l})-\frac{1}{2}\sum_{l,k=0}^K\alpha_l\alpha_k\sum_{t=1}^n(x_{t-l}'x_{t-k}'-x_{t-l}x_{t-k})\geq0}\nonumber\\
& = \left\{\bx'\in\mathbb{R}^n:\;\sum_{l=0}^K\alpha_l\sum_{t=1}^ny_t(x_{t-l}'-x_{t-l})-\frac{1}{2}\norm{\balpha}_2^2\sum_{t=1}^n(x_t'^{2}-x_t^2)\nonumber\right.\\
&\left.\ \ \ \ \ \ \ \ \ \ \ \ \ \ \ \ \ \ \ \ -\sum_{l=1}^K\Pi_l(\balpha)\sum_{t=1}^n(x_{t}'x_{t-l}'-x_{t}x_{t-l})\geq0\right\}\label{A2}\\
& = \ppp{\bx'\in\mathbb{R}^n:\;\psi(\bx',\by)-\psi(\bx,\by)\geq0}
\end{align}
where $\psi(\bx,\by)$ is given in \eqref{bpsivec}, and \eqref{A2} follows from the circularity assumption \eqref{circularAssump}, indeed,
\begin{align}
\sum_{l=0}^K\sum_{k=0}^K\alpha_l\alpha_k\sum_{t=1}^n x_{t-l}x_{t-k} &= \norm{\balpha}_2^2\sum_{t=1}^nx_t^2+2\cdot\sum_{l=0}^K\sum_{k=l+1}^K\alpha_l\alpha_k\sum_{t=1}^n x_{t}x_{t-(k-l)}\\
& = \norm{\balpha}_2^2\sum_{t=1}^nx_t^2+2\cdot\sum_{l=0}^K\sum_{m=1}^{K-l}\alpha_l\alpha_{m+l}\sum_{t=1}^n x_{t}x_{t-m}\\
& = \norm{\balpha}_2^2\sum_{t=1}^nx_t^2 + 2\cdot\sum_{l=1}^K\Pi_{l}(\balpha)\sum_{t=1}^n x_{t}x_{t-l}\label{circular2}
\end{align} 
where $\Pi_{l}(\balpha)$ is defined in \eqref{prodalphadef}. 

Then, we may write
\begin{align}
\int_{\calV(\bxt,\byt)}\mathrm{d}\bx'\mu_p(\bx') &= \int_{\mathbb{R}^n}\mathrm{d}\bx'\mu_p(\bx')\Ind\ppp{\psi(\bx',\by)-\psi(\bx,\by)\geq0}\\
& = \frac{1}{2\pi j}\int_{\mathbb{R}^n}\mathrm{d}\bx'\mu_p(\bx')\int_{c-j\infty}^{c+j\infty}\mathrm{d}\omega\frac{1}{\omega}\exp\ppp{\omega\p{\psi(\bx',\by)-\psi(\bx,\by)}}\\
& = \frac{1}{2\pi j}\int_{c-j\infty}^{c+j\infty}\mathrm{d}\omega\frac{1}{\omega}\exp\ppp{-\omega\psi(\bx,\by)}\int_{\mathbb{R}^n}\mathrm{d}\bx'\mu_p(\bx')\exp\ppp{\omega\psi(\bx',\by)}
\end{align}
where
\begin{align}
\mu_p(\bx) = \frac{1}{(2\pi\eta^2)^{n/2}}\exp\ppp{-\frac{1}{2\eta^2}\sum_{t=1}^n\p{x_t-\sum_{l=1}^p\varphi_lx_{t-l}}^2},
\end{align}
and due to the circularity assumption we may write
\begin{align}
\mu_p(\bx) &= \frac{1}{(2\pi\eta^2)^{n/2}}\exp\ppp{-\frac{1}{2\eta^2}\pp{\p{1+\norm{\bvarphi}_2^2}\norm{\bx}_2^2+2\sum_{l=1}^{p}(\hat{\Pi}_{l}(\bvarphi)-\varphi_l)\sum_{t=1}^nx_tx_{t-l}}}\\
& \triangleq \frac{1}{(2\pi\eta^2)^{n/2}}\exp\ppp{-\frac{1}{2\eta^2}L(\bx)}.
\end{align}
Thus,
\begin{align}
\int_{\calV(\bxt,\byt)}\mathrm{d}\bx'\mu_p(\bx') &= \frac{1}{2\pi j}\frac{1}{(2\pi\eta^2)^{n/2}}\int_{c-j\infty}^{c+j\infty}\mathrm{d}\omega\frac{\exp\ppp{-\omega\psi(\bx,\by)}}{\omega}\nonumber\\
&\ \ \ \ \ \ \ \ \ \ \ \ \ \ \ \ \times\int_{\mathbb{R}^n}\mathrm{d}\bx'\exp\ppp{-\frac{1}{2\eta^2}L(\bx')+\omega\psi(\bx',\by)}.\label{beforesaddle}
\end{align}
Now, the last exponent at the r.h.s. of the above equality can be rewritten as
\begin{align}
\frac{1}{2\eta^2}L(\bx)-\omega\psi(\bx',\by) &= \bx'^T\mathbf{\Omega}\bx'-\omega\by^T\mathbf{A}\bx'\\
& = (\bx'-\frac{\omega}{2}\mathbf{\Omega}^{-1}\mathbf{A}\by)^T\mathbf{\Omega}(\bx'-\frac{\omega}{2}\mathbf{\Omega}^{-1}\mathbf{A}\by)-\frac{\omega^2}{4}{\by}^T\mathbf{A}\mathbf{\Omega}^{-1}\mathbf{A}^T{\by}
\end{align}
where $\mathbf{\Omega}$ and $\mathbf{A}$ are defined in \eqref{omega1}-\eqref{omega2} and \eqref{Amat}, respectively. Then,
\begin{align}
&\int_{\mathbb{R}^n}\mathrm{d}\bx'\exp\ppp{-\frac{1}{2\eta^2}L(\bx')+\omega\psi(\bx',\by)} = \pp{\det\p{2\pi\cdot\frac{1}{2}\mathbf{\Omega}^{-1}}}^{1/2}\exp\ppp{\frac{\omega^2}{4}{\by}^T\mathbf{A}\mathbf{\Omega}^{-1}\mathbf{A}^T{\by}}\nonumber\\
&\hspace{5cm}= \exp\ppp{\frac{n}{2}\log\pi-\frac{1}{2}\log\det\mathbf{\Omega}+\frac{\omega^2}{4}{\by}^T\mathbf{A}\mathbf{\Omega}^{-1}\mathbf{A}^T{\by}}
\end{align}
where in the first equality we have used the fact that $\mathbf{\Omega}$ is a symmetric positive-definite Toeplitz matrix.\footnote{Indeed, from \eqref{omega1}-\eqref{omega2} we see that $\mathbf{\Omega}$ is diagonally dominant matrix with positive diagonal elements, and thus positive-definite.} Therefore, using the last result and \eqref{beforesaddle}, we get that
\begin{align}
\int_{\calV(\bxt,\byt)}\mathrm{d}\bx'\mu_p(\bx') &\doteq \frac{1}{(2\pi\eta^2)^{n/2}}\int_{c-j\infty}^{c+j\infty}\mathrm{d}\omega\frac{1}{\omega}\exp\pp{nq_n(\omega,\bx,\by)},\label{beforeLaplace}
\end{align}
and $q_n(\omega,\bx,\by)$ is defined in \eqref{hdef}. The integral in \eqref{beforeLaplace} can now be assessed using the saddle-point method \cite{Bruijn,NeriMono}. The derivative of $q_n(\omega,\bx,\by)$ vanishes at the value of $\omega$ that solves the equation $\partial q_n(\omega,\bx,\by)/\partial\omega=0$, where the gradient is taken w.r.t. $\omega$. We will show that this saddle-point, denoted by $\omega^*$, is in fact real-valued, i.e., $\omega^*\in\mathbb{R}_+$. Accordingly, we choose $c=\omega^*$, and thereby let the integration path pass through this saddle-point. Now, at $\omega^*$, $q_n(\omega,\bx,\by)$ has its maximum along the vertical direction $\omega = \omega^*+j\kappa, \;-\infty<\kappa<\infty$ (and hence it dominates the integral), but since it is a saddle-point it minimizes $q_n(\omega,\bx,\by)$ in the horizontal direction (the real line), so we get
\begin{align}
\int_{\calV(\bxt,\byt)}\mathrm{d}\bx'\mu_p(\bx') &\doteq  \frac{1}{(2\pi\eta^2)^{n/2}}\exp\ppp{n\cdot\min_{\omega}q_n(\omega,\bx,\by)}\label{1VolEstimate0}\\
&\triangleq\exp\ppp{n\cdot h_n(\bx,\by)}\label{VolEstimate0}
\end{align}
where $h_n(\bx,\by)$ is defined in \eqref{hndef}. 

We next show that $\omega^*\in\mathbb{R}_+$, as claimed above. Indeed, the modulus of the integrand in \eqref{beforeLaplace} depends solely on the real part of the exponent of the integrand, namely, on $\mathrm{Re}\ppp{q_n(\omega,\bx,\by)}$. Now, if we consider an arbitrary complex number $\omega = \omega_R+j\cdot\omega_I$, then we need to show that $\mathrm{Re}\ppp{q_n(\omega,\bx,\by)}$ is maximized only at $\omega_I=0$. Let $\bar{\mathbf{\Omega}}_1\triangleq 2\omega^{-1}\cdot\mathbf{\Omega}_1$, and then by definition $\mathbf{\Omega} = \omega\cdot\bar{\mathbf{\Omega}}_1+\mathbf{\Omega}_2$. Also, define $\mathbf{V}\triangleq (\omega_R\cdot\barOmega_1+\bOmega_2)^2+\omega_I^2\cdot\barOmega_1^2$, where for a symmetric matrix $\mathbf{A}$, by $\mathbf{A}^2$ we mean $\mathbf{A}\mathbf{A}^T$. Recall that
\begin{align}
q_n(\omega,\bx,\by)\triangleq \frac{1}{2}\log\pi-\frac{1}{2n}\log\det\mathbf{\Omega}+\frac{\omega^2}{4n}{\by}^T\mathbf{A}\mathbf{\Omega}^{-1}\mathbf{A}^T{\by}-\frac{\omega}{n}\psi(\bx,\by).\label{qdeft}
\end{align}
It is a simple exercise to check that $-\mathrm{Re}\ppp{\log\det\mathbf{\Omega}}$ is maximized at $\omega_I=0$, and since the real part of the first and last terms in \eqref{qdeft} are independent of $\omega_I$, we focus on the real part of the third term, which after simple algebra boils down to
\begin{align}
\mathrm{Re}\ppp{\frac{\omega^2}{4}{\by}^T\mathbf{A}\mathbf{\Omega}^{-1}\mathbf{A}^T{\by}} = \by^T\mathbf{A}\pp{\mathbf{V}^{-1}\p{\omega_R^3\barOmega_1+\omega_R^2\bOmega_2+\omega_R\omega_I^2\barOmega_1-\omega_I^2\bOmega_2}}\mathbf{A}^T\by.\label{realpart}
\end{align}
To prove that \eqref{realpart} is maximized at $\omega_I=0$ it suffices to show that
\begin{align}
\mathbf{V}^{-1}\p{\omega_R^3\barOmega_1+\omega_R^2\bOmega_2+\omega_R\omega_I^2\barOmega_1-\omega_I^2\bOmega_2}\preceq \pp{\p{\omega_R\cdot\barOmega_1+\bOmega_2}^2}^{-1}\p{\omega_R^3\barOmega_1+\omega_R^2\bOmega_2}\label{semi}
\end{align}
where for two matrices $\mathbf{X}$ and $\mathbf{Z}$, $\mathbf{X}\preceq\mathbf{Z}$ means that $\mathbf{Z}-\mathbf{X}$ is semi-positive definite matrix. Simple algebra reveals that to prove \eqref{semi} it suffices to show that
\begin{align}
\p{\omega_R^3\barOmega_1+\omega_R^2\bOmega_2}\barOmega_1^2\succeq\p{\omega_R\barOmega_1-\bOmega_2}\p{\omega_R\cdot\barOmega_1+\bOmega_2}^2,\label{expand}
\end{align}
which follows by expanding the terms on the left and right hand side of \eqref{expand}, and using the fact that $\barOmega_1$ and $\bOmega_2$ are semi-positive definite matrices. 

Finally, we show that the convergence in \eqref{1VolEstimate0} is uniform over $(\bx,\by)\in\mathcal{H}_n(B)$, as claimed in the statement of Lemma~\ref{lem:volume}. Precisely, we show that for any $\epsilon>0$, there exist an $N_\epsilon$ large enough and independent of $(\bx,\by)$, such that for any $n>N_\epsilon$
\begin{align}
\abs{\frac{1}{n}\log\int_{\calV(\bxt,\byt)}\mathrm{d}\bx'\mu_p(\bx')-h_n(\bx,\by)}<\epsilon.\label{uniformity}
\end{align}
Let $v(z)$ be an analytical function, and let $z_0\in\mathbb{R}$ be its unique saddle-point. By the analyticity of $v(\cdot)$, for any $\epsilon>0$, there exist a $\delta>0$ such that:
\begin{align}
\abs{v(z)-v(z_0)-\frac{1}{2}v''(z_0)(z-z_0)^2}\leq\epsilon(z-z_0)^2,\label{eq:first_cond}
\end{align}
for any $z$ such that $\abs{z-z_0}<\delta$. Accordingly, it can be shown that \cite{Bruijn}
\begin{align}
e^{n v(z_0)} \cdot  \left[\sqrt{\frac{2\pi}{|v''(z_0) + \epsilon|n}} -  2Q\left(\delta\sqrt{n\left(|v''(z_0)+\epsilon|\right)}\right)   \right] &\leq \int_{z_0 - j\infty}^{z_0 + j \infty} \exp nv(z) \mathrm{d}z \nonumber \\
&\leq e^{n v(z_0)}\sqrt{\frac{2\pi}{|v''(z_0) - \epsilon|n}} + C, \label{eq:lemma_saddle}
\end{align}
for some constant $C$, and $Q(x)\triangleq\int_x^\infty(2\pi)^{-1/2}e^{-t^2/2}\mathrm{d}t$, for $x\in\mathbb{R}$. To prove \eqref{uniformity}, we show that the terms in the square brackets at the left and right hand sides of \eqref{eq:lemma_saddle} can be lower and upper bounded by some universal constants. The proof consists of two steps. First, we will show that for any $\epsilon>0$, there exists a universal $\delta>0$ independent of $(\bx,\by)$, such that $q_n(\omega,\bx,\by)$ is close to its Taylor approximation around a ball of radius $\delta$ centered at the saddle-point. Then, we give a bound on $\abs{q_n''(\omega,\bx,\by)}$, which will be independent of $(\bx,\by)$. The desired result will then follow by using the last fact, $\abs{h''(z_0)\leq0}$, and \eqref{eq:lemma_saddle}.

Recall that $q_n(\omega,\bx,\by)$ converges almost surely to a value $f(\omega)$, as $n\to\infty$, for any $(\bx,\by) \in \mathcal{H}_n(B)$. Let $\omega_n^*(\bx,\by)$ be the sequence of saddle-points solving the minimization in \eqref{1VolEstimate0}. By the uniform continuity of $q_n(\omega,\bx,\by)$ w.r.t. $(\bx,\by)$, there is a radius $\Delta>0$ such that for $n$ sufficiently large, all saddle-points are contained in a ball centered at $0$ and of radius $\Delta$, i.e., $w_n^*(\bx,\by) \in \mathcal{B}(0,\Delta)$ almost surely. Next, we define
\begin{align}
M_\Delta  \triangleq \max_{(\bxt,\byt) \in \mathcal{H}_n(B)} \max_{|\omega - \omega_0|=\Delta} |q_n(\omega,\bx,\by)|. \label{eq:cauchy_est}
\end{align}
By the Cauchy integration formula around the circle of radius $\Delta$ above, with parametrization $\gamma$, we can obtain the following expression of the Taylor approximation error $E(\omega,\bx,\by)$ \cite[Section 3]{remmert2012theory}
\begin{align}
E(\omega,\bx,\by)&= q_n(\omega,\bx,\by) - q_n(w^*,\bx,\by) - \frac{1}{2}q_n''(w^*,\bx,\by) (\omega - w^*)^2 \\
&= \sum_{k=3}^\infty \frac{(\omega - \omega^*)^k}{2\pi j} \oint_\gamma \frac{q_n(z,\bx,\by)}{(z - \omega^*)^{j+1}}\mathrm{d}z
\label{eq:close_form_residual}
\end{align}
where $\omega^*=\omega_n^*(\bx,\by)$. Now, let us consider a ball of radius $\delta<\Delta$ around the saddle-point, such that $\mathcal{B}(\omega^*,\delta)\subset\mathcal{B}(0,\Delta)$. Using \eqref{eq:cauchy_est} and \eqref{eq:close_form_residual}, for any $\omega \in \mathcal{B}(\omega^*,\delta)$ we can bound $\abs{R(\omega,\bx,\by)}$ as follows
\begin{align}
\abs{E(\omega,\bx,\by)} &\leq \sum_{k = 3}^\infty \frac{M_\Delta |\omega - \omega^*|^k}{\Delta^k} \\
& \leq M_\Delta \frac{\delta^3}{\Delta^2(\Delta - \delta)} \triangleq \epsilon,
\end{align}
where the last step follows by evaluating the power series. Thus, for any $\epsilon>0$, there exist a $\delta$ independent of $(\bx,\by)$ such that \eqref{eq:first_cond} is verified. Finally, by using Cauchy integration again, we have
\begin{align}
|q_n''(\omega,\bx,\by)| &= \frac{1}{\pi} \oint_\gamma \frac{q_n(z,\bx,\by)}{(z - \omega)^2} \\
& \leq \frac{2M_\Delta}{\Delta^2},\label{uniforIssue}
\end{align}
as required.

\section{Proof of Theorem~\ref{th:2}}\label{sec:proof2}

In this section, we prove Theorem~\ref{th:2}. Some technical details will be omitted, since they follow from similar steps used in the proof of Theorem~\ref{th:1}. Fix an arbitrary $\epsilon>0$, pick $p\in\mathbb{N}$, and a set of parameters $\ppp{\gamma_k}_{k=0}^p\in\Gamma$. The codebook $\mathfrak{C}_n=\ppp{\bx_i}_{i=1}^M$ is generated by drawing $M$ codewords independently and uniformly at random over $\mathcal{T}^n_\varepsilon(\bgamma)\triangleq\mathcal{T}^n_\varepsilon$, defined in \eqref{condtypemar}. Without loss of generality we assume that the codeword $\bx\triangleq\bx_1$ was sent. Then, 
similarly as in \eqref{barPenRSurvive}-\eqref{lowUppermar}, we have
\begin{align}
\bar{P}_{e}(n,R) &\doteq\bE\pp{\calI\ppp{\calH_n(B)}\cdot\min\ppp{1,M\cdot\Pr\ppp{\left.\frac{1}{n}\log V\p{\bY\vert \bX_2}\geq \frac{1}{n}\log V\p{\bY\vert \bX_1}\right\vert\calF_0}}}.\label{barPenRSurvive2}
\end{align}
For $(\bX_1,\bY) = (\bx,\by)\in\calH_n(B)$, the inner probability term at the r.h.s. of \eqref{barPenRSurvive2} can be represented as
\begin{align}
\Pr\ppp{\frac{1}{n}\log V\p{\by\vert \bX_2}\geq \frac{1}{n}\log V\p{\by\vert \bx}} &= \frac{1}{\mathrm{Vol}(\mathcal{T}_\varepsilon^n)}\int_{\bxt'\in\calV(\bxt,\byt)\cap\mathcal{T}^n_\varepsilon} \mathrm{d}\bx'\\
& = \frac{\mathrm{Vol}\left(\mathcal{V}(\bx,\by) \cap \mathcal{T}_\varepsilon^n\right)}{\mathrm{Vol}(\mathcal{T}_\varepsilon^n)}\label{lowUppermar2}
\end{align}
where
\begin{align}
\mathcal{V}(\bx,\by) \cap \mathcal{T}_\varepsilon^n &\triangleq \ppp{\bx' \in \mathcal{T}_\varepsilon^n:\;\frac{1}{n}\log V\p{\by\vert \bx'}\geq \frac{1}{n}\log V\p{\by\vert \bx}} \\
& = \left\{\bx'\in \mathcal{T}_\varepsilon^n:\;\sum_{l=0}^K\alpha_l\sum_{t=1}^ny_t(x_{t-l}'-x_{t-l})-\frac{1}{2}\norm{\balpha}_2^2\sum_{t=1}^n(x_t'^{2}-x_t^2)\nonumber\right.\\
&\left.\hspace{3cm} -\sum_{l=1}^K\Pi_l(\balpha)\sum_{t=1}^n(x_{t}'x_{t-l}'-x_{t}x_{t-l})\geq0\right\}. \label{eq:error_event}
\end{align}
Due to the fact that $\bx',\bx\in \mathcal{T}_\varepsilon^n$,
\begin{align}
\abs{\sum_{t=1}^n (x_t'x_{t-l}' - x_t x_{t-l})}\leq 2n\varepsilon,\ l\in[0:p],\label{trueforT}
\end{align}
and therefore,
\begin{align}
\mathcal{V}^{-}(\bx,\by) \cap \mathcal{T}_\varepsilon^n \subseteq \mathcal{V}(\bx,\by) \cap\mathcal{T}_\varepsilon^n \subseteq \mathcal{V}^+(\bx,\by) \cap\mathcal{T}_\varepsilon^n,\label{eq:bounds}
\end{align}
in which 
\begin{align}
\mathcal{V}^+(\bx,\by) &\triangleq \left\{\bx' \in \mathbb{R}^n : \; \sum_{l=0}^K\alpha_l\sum_{t=1}^ny_t(x_{t-l}'-x_{t-l})\right.\nonumber\\
&\left. \ \ \ \ \ \ \ \ \ -\sum_{l=1+p\wedge K}^K\Pi_l(\balpha)\sum_{t=1}^n(x_{t}'x_{t-l}'-x_{t}x_{t-l}) + \norm{\balpha}_2^2 n\varepsilon + 2n \varepsilon\sum_{l=1}^K\Pi_l(\balpha)\geq0\right\}\nonumber\\
&\triangleq \ppp{\bx' \in \mathbb{R}^n : \; \psi'(\bx', \by) - \psi'(\bx,\by) + nC^+(\balpha)\varepsilon \geq 0}\label{V+}
\end{align}
where 
\begin{align}
\psi'(\bx,\by) &\triangleq \sum_{l=0}^K \alpha_l \sum_{t=1}^n y_t x_{t-l}-\sum_{l=1+p\wedge K}^K\Pi_l(\balpha)\sum_{t=1}^nx_{t}x_{t-l}\\
C^+(\balpha) &\triangleq \| \balpha\|^2 + 2\sum_{l=1}^K \Pi_l(\balpha),
\end{align} 
and $\mathcal{V}^-(\bx,\by)$ is defined similarly to \eqref{V+} but with $C^+(\balpha)$ replaced by $C^-(\balpha) \triangleq -C^+(\balpha)$. 

The bulk of the argument resides in understanding the exponential behavior of \eqref{lowUppermar2}. Using \eqref{eq:bounds}, we note that
\begin{align}
\mathrm{Vol}\left(\mathcal{V}^-(\bx,\by) \cap \mathcal{T}_\varepsilon^n\right) \leq \mathrm{Vol}\left(\mathcal{V}(\bx,\by) \cap\mathcal{T}_\varepsilon^n\right) \leq \mathrm{Vol}\left(\mathcal{V}^+(\bx,\by) \cap\mathcal{T}_\varepsilon^n\right).
\end{align}
In the sequel, we study the upper bound, with the understanding that the lower bound follows from similar steps. We write
\begin{align}
\mathrm{Vol}\left(\mathcal{V}^+(\bx,\by)\cap\mathcal{T}_\varepsilon^n\right) &= \int_{\bx'\in\mathbb{R}^n}\Ind\ppp{\psi'(\bx',\by) - \psi'(\bx,\by) + n\varepsilon C^+(\balpha) \geq 0} \nonumber \\ 
&  \ \ \ \ \ \ \ \ \ \cdot\prod_{k=0}^p\Ind\ppp{\abs{\frac{1}{n}\sum_{t=1}^n x'_t x'_{t-k} - \gamma_k} \leq \varepsilon} \mathrm{d}\bx'.\label{sad1}
\end{align}
Recalling that $\Ind \ppp{|a| < \varepsilon} = \left[ \Ind \ppp{a \leq \varepsilon} - \Ind \ppp{a \leq - \varepsilon} \right]$, and using \eqref{eq:laplace}, we get
\begin{align}
\mathrm{Vol}\left(\mathcal{V}^+(\bx,\by) \cap \mathcal{T}_\varepsilon^n\right)&\doteq \int_{\bx' \in \mathbb{R}^n} \mathrm{d}\bx'  \int_{c_{p+1} - j\infty}^{c_{p+1} + j \infty}\mathrm{d}\omega_{p+1} \frac{\exp \{ \omega_{p+1}( \psi'(\bx',\by) - \psi'(\bx,\by) + n\varepsilon C^+(\balpha)) \}}{\omega_{p+1}} \nonumber \\
&\ \ \ \ \cdot \prod_{k=0}^p \int_{c_k - j\infty}^{c_k + j \infty} \mathrm{d}\omega_k \exp \left\{ \omega_k \left( n\gamma_k - \sum_{t = 1}^n x'_t x'_{t-k}  \right)\right\} \left( \frac{e^{n \omega_k \varepsilon} - e^{-n \omega_k \varepsilon}}{w_k}  \right). 
\end{align}
Exchanging the order of integrals and collecting terms, we obtain
\begin{align}
\mathrm{Vol}\p{\mathcal{V}^+(\bx,\by) \cap \mathcal{T}_\varepsilon^n} &\doteq\int_{c_{p+1} - j \infty}^{c_{p+1} + j\infty} \mathrm{d}\omega_{p+1} \frac{1}{\omega_{p+1}} \exp \{ -\omega_{p+1}\psi'(\bx,\by) + n\varepsilon C^+(\balpha)\} \nonumber \\
& \hspace{1em} \cdot \int_{\bar{\mathcal{W}}} \mathrm{d}\bomega \left[ \prod_{k=0}^p \left( \frac{e^{n \omega_k \varepsilon} - e^{-n \omega_k \varepsilon}}{w_k}  \right) \exp\left\{ n \omega_k  \gamma_k \right\} \right] \nonumber \\
& \hspace{1em} \cdot \int_{\bx' \in \mathbb{R}^n} \mathrm{d}\bx' \exp\{\omega_{p+1}\psi'(\bx',\by)\}\prod_{k=0}^p \exp\left\{ -\omega_k \sum_{t=1}^n x'_t x_{t-k}'  \right\},\label{sad2}
\end{align}
where $\bar{\mathcal{W}}\triangleq\{ \bomega : c_k - j\infty \leq \omega_k \leq c_k + j\infty,\; k = 0,\ldots,p \}$. Next, we note that
\begin{align}
\omega_{p+1} \psi'(\bx',\by) - \sum_{l=0}^p \omega_l \sum_{t=1}^n x_t' x'_{t-l} & = \omega_{p+1} \sum_{l=0}^K \alpha_l \sum_{t=1}^n y_t x'_{t-l}-\omega_{p+1}\cdot\sum_{l=1+p\wedge K}^K\Pi_l(\balpha)\sum_{t=1}^nx_{t}'x_{t-l}'\nonumber\\
& \ \ \ \ - \sum_{l=0}^p \omega_l \sum_{t=1}^n x_t' x'_{t-l} \\
& = \omega_{p+1} \cdot \by^T \bA \bx' - \bx'^T \mathbf{\Omega}_0 \bx' \\
& = - \left(\bx' + \frac{\omega_{p+1}}{2}\mathbf{\Omega}_0^{-1} \bA^T \by \right)^T\mathbf{\Omega}_0\left(\bx' + \frac{\omega_{p+1}}{2}\mathbf{\Omega}_0^{-1}\bA^T \by \right)  \nonumber \\ 
& \hspace{1em} +  \frac{\omega_{p+1}^2}{4} \by^T \bA \mathbf{\Omega}_0^{-1} \bA^T \by,
\end{align}
where the last step follows by completing the square, and $\mathbf{\Omega}_0$ is the $n\times n$ symmetric Toeplitz matrix,
\begin{align}
\pp{\mathbf{\Omega}_0}_{i,j} \triangleq\begin{cases}
\omega_0,&\ \text{for}\;\abs{i-j}=0\\
0.5\cdot\omega_{\abs{i-j}},&\ \text{for}\;1\leq\abs{i-j}\leq p\\
0.5\cdot\omega_{p+1}\cdot\Pi_{\abs{i-j}}(\balpha),&\ \text{for}\;1\leq p\wedge K\leq\abs{i-j}\leq K\\
0, &\ \text{otherwise}
\end{cases}.\label{omega0}
\end{align}
Therefore,
\begin{align}
&\int_{\bx' \in \mathbb{R}^n}\mathrm{d}\bx' \exp\{\omega_{p+1}\psi(\bx',\by)\}\prod_{l=0}^p \exp\left\{ -\omega_l \sum_{t=1}^n x'_t x_{t-l}'  \right\}\nonumber\\
&\ \ \ \ \ \ \ = \exp\left\{ \frac{n}{2}\log \pi - \frac{1}{2} \log \det \mathbf{\Omega}_0 + \frac{\omega_{p+1}^2}{4} \by^T \bA \mathbf{\Omega}_0^{-1} \bA^T \by \right\} \nonumber
\end{align}
where in the last equality we have used the fact that $\mathbf{\Omega}_0$ is a symmetric positive-definite Toeplitz matrix which follows because $\bomega\in\calW$. Using the last result, we finally get
\begin{align}
\mathrm{Vol}\left(\mathcal{V}^+(\bx,\by) \cap \mathcal{T}_\varepsilon^n\right) \doteq  \int_{\mathcal{W}} \mathrm{d}\bomega \exp \{ n \bar{q}_n(\bomega,\bx,\by,\varepsilon)\}, \label{eq:saddle_point_int}
\end{align}
where 
\begin{align}
\bar{q}_n(\bomega,\bx,\by,\varepsilon) &= \frac{\log \pi}{2} - \frac{1}{2n} \log \det \mathbf{\Omega}_0 + \frac{\omega_{p+1}^2}{4n} \by^T \bA \mathbf{\Omega}_0^{-1} \bA^T \by - \frac{\omega_{p+1}}{n}\psi'(\bx,\by) + \sum_{l= 0}^p \omega_l \gamma_l  \nonumber \\ &  \hspace{1em} + \frac{1}{n} \sum_{l=0}^p \log \sinh(n \omega_l \varepsilon) + \varepsilon C^+(\balpha).
\end{align}
Using similar steps as in Appendix~\ref{app:1}, the integral in \eqref{eq:saddle_point_int} can be evaluated using the saddle-point method, resulting in
\begin{align}
\mathrm{Vol}\left(\mathcal{V}^+(\bx,\by) \cap \mathcal{T}_\varepsilon^n\right) \doteq \exp\ppp{n\bar{h}_n(\bx,\by,\varepsilon)},\label{resgenfixedens}
\end{align}
where $\bar{h}_n(\bx,\by,\varepsilon) = \min_{\bomegat\in\calW} q_n(\bomega,\varepsilon)$. Also, using the same steps as in \eqref{VolEstimate0}-\eqref{uniforIssue}, one can show that the saddle-point solution is in fact real-valued vector, and that the convergence \eqref{resgenfixedens} is uniform w.r.t. $(\bx,\by)$. Similarly, one can verify that
\begin{align}
\lim_{\varepsilon \to 0}\lim_{n\to\infty} \frac{1}{n}\log\mathrm{Vol}(\mathcal{T}_\varepsilon^n)= \frac{1}{2} \log \left( 2 \pi e  \eta^2 \right).\label{eq:entr}
\end{align}
Thus, we may conclude that
\begin{align}
\bar{P}_{e}(n,R) \doteq \bE\pp{\calI\ppp{\calH_n(B)}\exp\ppp{-n\pp{\frac{1}{2} \log \left( 2 \pi e  \eta^2 \right)-\bar{h}_n(\bX,\bY,\varepsilon)-R+o(\varepsilon)}_+}}.\label{probaafter}
\end{align}
The next step is taking the expectation w.r.t. $(\bX,\bY)$, distributed according to $\mu_p \times W$. Similarly as in the proof of Theorem~\ref{th:1}, we calculate the limit of each term in $\bar{q}_n(\bomega,\bx,\by,\varepsilon)$. By Szeg\"{o} theorem,
\begin{align}
\frac{1}{n} \log \det \mathbf{\Omega}_0 =  \frac{1}{2\pi} \int_{0}^{2\pi} \log g_{\bomegat}(\nu) d\nu,
\end{align}
where $g_{\bomegat}(\cdot)$ is defined in \eqref{gomegadef}. Using the same steps as in \eqref{LLNa}-\eqref{LLNa2}, by the LLN and Szeg\"{o} theorem, we get with overwhelming probability as $n\to\infty$,
\begin{align}
\lim_{n\to\infty}\frac{1}{n}\by^T\mathbf{A}\mathbf{\Omega}_0^{-1}\mathbf{A}^T\by &= \lim_{n\to\infty}\frac{1}{n}\bE\ppp{\bY^T\mathbf{A}\mathbf{\Omega}_0^{-1}\mathbf{A}^T\bY}\\
& = \lim_{n\to\infty}\frac{1}{n}\tr\ppp{\mathbf{A}\mathbf{\Omega}_0^{-1}\mathbf{A}^T\mathbf{R}_{\byt}}\\
& = \frac{1}{2\pi}\int_{\pp{0,2\pi}}\mathrm{d}\nu\frac{S_Y(\nu)\abs{A(\nu)}^2}{g_{\bomegat}(\nu)},
\end{align}
and
\begin{align}
\lim_{n \to \infty} \frac{1}{n} \psi'(\bx,\by) = \sum_{l=0}^K \sum_{i=0}^K \alpha_l h_i \gamma_{l-i}-\sum_{l=1+p\wedge K}^K\Pi_l(\balpha)\gamma_l.
\end{align}
Also, recalling the fact that $\lim_{x\to\infty} (\log \sinh(x))/x = 1$,
\begin{align}
\lim_{n \to \infty} \frac{1}{n} \sum_{l=0}^p \log \sinh (n \omega_l \varepsilon) = \varepsilon \sum_{l = 0}^p \omega_l.
\end{align}
Collecting the last result, with probability approaching 1 as $n\to\infty$,
\begin{align}
\lim_{\varepsilon \to 0} \lim_{n \to \infty}\bar{h}_n(\bX,\bY,\varepsilon) &= \min_{\bomegat\in\Gamma}\left\{ \frac{\log \pi}{2} -\frac{1}{4\pi} \int_{0}^{2\pi} \log g_{\bomegat}(\nu) d\nu + \frac{\omega_{p+1}^2}{8\pi}\int_{\pp{0,2\pi}}\mathrm{d}\nu\frac{S_Y(\nu)\abs{A(\nu)}^2}{g_{\bomegat}(\nu)} \right.\nonumber\\
&\ \ \ \ \ \ \ \ \left.\vphantom{\int_{\pp{0,2\pi}}\mathrm{d}\nu\frac{S_Y(\nu)\abs{A(\nu)}^2}{g_{\bomegat}(\nu)}}-\omega_{p+1}\pp{\sum_{l=0}^K \sum_{i=0}^K \alpha_l h_i \gamma_{l-i}-\sum_{l=1+p\wedge K}^K\Pi_l(\balpha)\gamma_l} + \sum_{k= 0}^p \omega_k\gamma_k\right\}\\
&\triangleq -\hat{I}_2(\balpha,\bgamma). \label{eq:cond_entr}
\end{align}

Next, for any $\delta>0$, define the set
\begin{align}
\calA_n(\delta)&\triangleq\ppp{(\bx,\by):\;\abs{\bar{h}_n(\bx,\by,\varepsilon)+\hat{I}_2(\balpha,\bgamma)}<\delta}\label{Adef2}.
\end{align}
Accordingly, using \eqref{probaafter} we have
\begin{align}
\bar{P}_{e}(n,R) &\leq \bE\pp{\calI\ppp{\calA_n(\delta)}\exp\ppp{-n\pp{\frac{1}{2} \log \left( 2 \pi e  \eta^2 \right)-\bar{h}_n(\bX,\bY,\varepsilon)-R+o(\varepsilon)}_+}}+\Pr\ppp{\calA_n^c(\delta)},
\label{probaafter2}
\end{align}
and in light of \eqref{eq:cond_entr}, taking $\delta\to0$ and $\varepsilon\to0$ followed by $n\to\infty$, the last term at the r.h.s. of \eqref{probaafter2} is asymptotically negligible. Hence, in terms of achievable rate, we get that $\bar{P}_{e}(n,R)$ decays to zero as long as
\begin{align}
R&<\frac{1}{2} \log \left( 2 \pi e  \eta^2 \right)+\hat{I}_2(\balpha,\bgamma)=\bar{I}_2(\balpha,\bgamma).
\end{align}
Finally, since $\ppp{\gamma_k}_{k=1}^p\in\Gamma$ were fixed parameters up to this point, we take the maximum of the r.h.s. of the above inequality over these parameters, which concludes the proof.

\section{Proof of Theorem~\ref{th:3}}\label{app:3}
We derive the error exponent in the case where $\alpha_k=0$ for $1\leq k\leq K$, using the ensemble in \eqref{condtypemar} with $p=0$. Some technical details will be omitted, as they follow from similar steps used in the proof of Theorem~\ref{th:1}. We use \eqref{probaafter} in Appendix~\ref{sec:proof2}, and we note that in the above special case \eqref{probaafter} simplifies to
\begin{align}
\bar{P}_{e}(n,R) \doteq \bE\pp{\calI\ppp{\calH_n(B)}\exp\ppp{-n\pp{I(\hat{\rho}_{XY},h_0,\alpha_0)-R+o(\varepsilon)}_+}}\label{peexpo}
\end{align}
where $\hat{\rho}_{XY}$ is the empirical correlation coefficient defined as
\begin{align}
\hat{\rho}_{XY}\triangleq\frac{\frac{1}{n}\sum_{t=1}^nX_tY_t}{\sqrt{P_X}\sqrt{\frac{1}{n}\sum_{t=1}^nY_t^2}},
\end{align}
and $I(\hat{\rho}_{XY},h_0,\alpha_0)$ is defined in \eqref{IdefExpo}. The next step is taking the expectation w.r.t. $(\bX,\bY)$ distributed according to $\mu\times W$. To this end, let $\upsilon>0$, and define the set
\begin{align}
\calL_{\upsilon,\varepsilon}(P_X,P_Y,\rho_{XY})&\triangleq\left\{\bx,\by\in\mathbb{R}^n:\;\abs{\frac{1}{n}\sum_{t=1}^nx_t^2-P_X}<\varepsilon,\;\abs{\frac{1}{n}\sum_{t=1}^ny_t^2-P_Y}<\upsilon,\right.\nonumber\\
&\left.\hspace{2.55cm}\abs{\frac{1}{n}\sum_{t=1}^nx_ty_t-\rho_{XY}\sqrt{P_XP_Y}}<\upsilon\right\}.
\end{align}
With this definition we see that the exponent term in \eqref{peexpo} is almost constant over $\calL_{\upsilon,\varepsilon}(P_X,P_Y,\rho_{XY})$, namely, $\abs{I(\hat{\rho}_{XY})-I(\rho_{XY})}<\xi(\upsilon)$ with $\xi(\upsilon)\to0$ as $\upsilon\to0$. Then, we may write
\begin{align}
\bar{P}_{e}(n,R)&\doteq \int_{\calH_n(B)}\mathrm{d}\bx\mathrm{d}\by \exp\ppp{-n\pp{I(\hat{\rho}_{XY})-R+o(\varepsilon)}_+}\mu(\bx)W(\by\vert\bx)\nonumber\\
&\doteq\int_{P_Y\in\pp{0,B},\abs{\rho_{XY}}\leq1}\mathrm{d}P_Y\mathrm{d}\rho_{XY}\int_{\calH_n(B)\cap\calL_{\upsilon,\varepsilon}(P_X,P_Y,\rho_{XY})}\hspace{-2cm}\mathrm{d}\bx\mathrm{d}\by \exp\ppp{-n\pp{I(\hat{\rho}_{XY})-R+o(\varepsilon)}_+}\mu(\bx)W(\by\vert\bx)\nonumber\\
&\doteq\max_{P_Y,\rho_{XY}}\int_{\calH_n(B)\cap\calL_{\upsilon,\varepsilon}(P_X,P_Y,\rho_{XY})}\mathrm{d}\bx\mathrm{d}\by \exp\ppp{-n\pp{I(\hat{\rho}_{XY})-R+o(\varepsilon)}_+}\mu(\bx)W(\by\vert\bx)\nonumber\\
&\doteq\max_{P_Y,\rho_{XY}}\exp\ppp{-n\pp{I(\rho_{XY})+\xi(\upsilon)-R+o(\varepsilon)}_+-\frac{n}{2}\log(2\pi eP_X)}\nonumber\\
&\hspace{2.5cm}\times\int_{\calH_n(B)\cap\calL_{\upsilon,\varepsilon}(P_X,P_Y,\rho_{XY})}\mathrm{d}\bx\mathrm{d}\by W(\by\vert\bx)\label{saddagain}
\end{align}
where the third (asymptotic) equality follows from the Laplace integration method. We next evaluate the integral term at the r.h.s. of \eqref{saddagain} using the saddle-point method as in \eqref{sad1}-\eqref{sad2}. We get
\begin{align}
\int_{\calH_n(B)\cap\calL_{\upsilon,\varepsilon}}\mathrm{d}\bx\mathrm{d}\by W(\by\vert\bx)&\doteq\int_{\hat{\calW}}\mathrm{d}\hat{\bomega} \exp\ppp{n\pp{\hat{\omega}_0P_X+\hat{\omega}_1\rho_{XY}\sqrt{P_XP_Y}+\hat{\omega}_2P_Y+o(\varepsilon)+o(\upsilon)}}\nonumber\\
&\hspace{0cm}\times\int_{\mathbb{R}^n\times\mathbb{R}^n}\mathrm{d}\bx\mathrm{d}\by W(\by\vert\bx)\exp\ppp{-\hat{\omega}_0\norm{\bx}_2^2-\hat{\omega}_1\sum_{t=1}^nx_ty_t-\hat{\omega}_2\norm{\by}_2^2}.
\end{align}
Define
\begin{align}
\pp{\mathbf{H}_1}_{i,j} \triangleq\begin{cases}
\hat{\omega}_0+\frac{1}{2}\norm{\bh}_2^2,&\ \text{for}\;\abs{i-j}=0\\
0.5\cdot\Pi_{\abs{i-j}}(\bh),&\ \text{for}\;1\leq\abs{i-j}\leq K\\
0, &\ \text{otherwise}
\end{cases}.\label{omega3}
\end{align} 
and let $\mathbf{H}_2$ be an $n\times n$ lower-triangular Toeplitz matrix where for $i\geq j$,
\begin{align}
\pp{\mathbf{H}_2}_{i,j} \triangleq\begin{cases}
\hat{\omega}_1-h_0,&\ \text{for}\;0\leq i-j=0\\
h_{i-j},&\ \text{for}\;1\leq i-j\leq K\\
0, &\ \text{otherwise}
\end{cases}.\label{Amat2}
\end{align}
Then, we have
\begin{align}
\log W(\by\vert\bx)&-\hat{\omega}_0\norm{\bx}_2^2-\hat{\omega}_1\sum_{t=1}^nx_ty_t-\hat{\omega}_2\norm{\by}_2^2 = -\frac{n}{2}\log(2\pi\sigma^2)-\frac{1}{2\sigma^2}\norm{\by}_2^2+\sum_{k=0}^Kh_k\sum_{t=1}^ny_tx_{t-k}\nonumber\\
&-\frac{1}{2}\norm{\bh}_2^2\norm{\bx}_2^2-\sum_{k=1}^K\Pi_k(\bh)\sum_{t=1}^nx_tx_{t-k}-\hat{\omega}_0\norm{\bx}_2^2-\hat{\omega}_1\sum_{t=1}^nx_ty_t-\hat{\omega}_2\norm{\by}^2\\
& = -\frac{n}{2}\log(2\pi\sigma^2)-\p{\frac{1}{2\sigma^2}+\hat{\omega}_2}\by^T\by-\bx^T\mathbf{H}_1\bx-\by^T\mathbf{H}_2\bx\\
& = -\frac{n}{2}\log(2\pi\sigma^2)-\pp{\lambda\norm{\by+\frac{1}{2\lambda}\mathbf{H}_2\bx}_2^2+\bx^T\p{\mathbf{H}_1-\frac{1}{4\lambda}\mathbf{H}_2^T\mathbf{H}_2}\bx}
\end{align}
where $\lambda\triangleq \frac{1}{2\sigma^2}+\hat{\omega}_2$. Therefore,
\begin{align}
&\int_{\mathbb{R^n}\times\mathbb{R}^n}\mathrm{d}\bx\mathrm{d}\by W(\by\vert\bx)\exp\ppp{-\hat{\omega}_0\norm{\bx}_2^2-\hat{\omega}_1\sum_{t=1}^nx_ty_t-\hat{\omega}_2\norm{\by}_2^2}\nonumber\\
&\hspace{1cm}= \exp\ppp{-\frac{n}{2}\log(2\pi\sigma^2)+n\log\pi-\frac{1}{2}\log\det\p{\frac{1}{4}\mathbf{H}_2^T\mathbf{H}_2-\lambda\cdot\mathbf{H}_1}}
\end{align}
where the last equality follows because $\hat{\bomega}\in\hat{\calW}$. Using the last result, we finally get
\begin{align}
\int_{\calH_n(B)\cap\calL_{\upsilon,\varepsilon}}\mathrm{d}\bx\mathrm{d}\by W(\by\vert\bx)&\doteq  \int_{\mathcal{W}} \mathrm{d}\hat{\bomega} \exp \ppp{ n\pp{ \tilde{q}_n(\hat{\bomega},\upsilon)-\frac{1}{2}\log(2\pi\sigma^2)+\log\pi+o(\varepsilon)+o(\upsilon)}}, \label{eq:saddle_point_intexpone}
\end{align}
where 
\begin{align}
\tilde{q}_n(\hat{\bomega},P_Y,\rho_{XY},\upsilon) &= \hat{\omega}_0P_X+\hat{\omega}_1\rho_{XY}\sqrt{P_XP_Y}+\hat{\omega}_2P_Y-\frac{1}{2n}\log\det\p{\frac{1}{4}\mathbf{H}_2^T\mathbf{H}_2-\lambda\cdot\mathbf{H}_1}.
\end{align}
Using similar steps as in Appendix~\ref{app:1}, the integral in \eqref{eq:saddle_point_int} can be evaluated using the saddle-point method, resulting in
\begin{align}
\int_{\calH_n(B)\cap\calL_{\upsilon,\varepsilon}}\mathrm{d}\bx\mathrm{d}\by W(\by\vert\bx)&\doteq \exp\ppp{n\pp{\tilde{h}_n(P_Y,\rho_{XY},\upsilon)-\frac{1}{2}\log(2\pi\sigma^2)+\log\pi+o(\varepsilon)+o(\upsilon)}},\label{resgenfixedensexp}
\end{align}
where $\tilde{h}_n(P_Y,\rho_{XY},\upsilon) = \min_{\hat{\bomegat}\in\hat{\calW}} \tilde{q}_n(\hat{\bomega},P_Y,\rho_{XY},\upsilon)$. Also, using the same steps as in \eqref{VolEstimate0}-\eqref{uniforIssue}, one can show that the saddle-point solution is in fact real-valued vector. Substituting the last result in \eqref{saddagain}, we get
\begin{align}
\bar{P}_{e}(n,R)&\doteq\max_{P_Y,\rho_{XY}}\exp\ppp{-n\pp{I(\rho_{XY})+\xi(\upsilon)-R+o(\varepsilon)}_+-\frac{n}{2}\log(2\pi eP_X)}\nonumber\\
&\ \ \ \times\exp\ppp{n\pp{\tilde{h}_n(P_Y,\rho_{XY},\upsilon)-\frac{1}{2}\log(2\pi\sigma^2)+\log\pi+o(\varepsilon)+o(\upsilon)}}\\
& = \max_{P_Y,\rho_{XY}}\exp\left\{\vphantom{\log\det\p{\frac{1}{4}\mathbf{H}_2^T\mathbf{H}_2-\lambda\cdot\mathbf{H}_1}}-n\pp{I(\rho_{XY})+\xi(\upsilon)-R+o(\varepsilon)}_+-\frac{n}{2}\log(4eP_X\sigma^2)\right.\nonumber\\
&\left.\hspace{4cm}+n\tilde{h}_n(P_Y,\rho_{XY},\upsilon)\right\}.\label{beffd0}
\end{align}
Now, using Szeg\"{o} theorem,
\begin{align}
\lim_{n\to\infty}\frac{1}{n} \log \det \p{\frac{1}{4}\mathbf{H}_2^T\mathbf{H}_2-\lambda\cdot\mathbf{H}_1} =  \frac{1}{2\pi} \int_{0}^{2\pi} \log u_{\hat{\bomegat}}(\nu) d\nu,
\end{align}
where $u_{\hat{\bomegat}}(\cdot)$ is defined in \eqref{uomegadef}, and so,
\begin{align}
\lim_{\upsilon\to0}\lim_{n\to\infty}\tilde{h}_n(P_Y,\rho_{XY},\upsilon) &= \min_{\hat{\bomegat}\in\hat{\calW}}\ppp{\hat{\omega}_0P_X+\hat{\omega}_1\rho_{XY}\sqrt{P_XP_Y}+\hat{\omega}_2P_Y-\frac{1}{4\pi} \int_{0}^{2\pi} \log u_{\hat{\bomegat}}(\nu) d\nu}\nonumber\\
&\triangleq-V(P_Y,\rho_{XY}).\label{beffd}
\end{align}
Using \eqref{beffd0} and \eqref{beffd}, we finally get
\begin{align}
\lim_{\upsilon,\varepsilon\to0}\lim_{n\to\infty}-\frac{1}{n}\log\bar{P}_{e}(n,R) = \min_{P_Y,\rho_{XY}}\ppp{V(P_Y,\rho_{XY})+\frac{1}{2}\log(4eP_X\sigma^2)+\pp{I(\rho_{XY})-R}_+}.
\end{align} 

\section{Proof of Theorem~\ref{th:4}}\label{app:4}

Due to similarities to the proofs of Theorems~\ref{th:1} and \ref{th:2}, we provide only a proof sketch. Fix an arbitrary $\varepsilon>0$. Define the sequence of sets $\mathcal{T}^n_\varepsilon(P_X)$, for $n=1,2,\ldots$, as follows
\begin{align}
\mathcal{T}^n_\varepsilon(P_X) &\triangleq \left\{\bx\in\mathbb{R}^n:\;\abs{\frac{1}{n}\sum_{t=1}^n x_t^2-P_X}<\varepsilon\right\}.
\end{align}
The codebook $\mathfrak{C}_n$ is generated by drawing $M$ codewords independently and uniformly at random from $\mathcal{T}^n_\varepsilon(P_X)$. Then, the probability of error corresponding to the universal decoder in \eqref{GLRT} is given by
\begin{align}
\bar{P}_{e}(n,R) &\doteq\bE\pp{\calI\ppp{\calH_n(B)}\cdot\min\ppp{1,M\cdot\Pr\ppp{\left.\abs{\bX_2^T\bY}\geq \abs{\bX_1^T\bY}\right\vert\calF_0}}}.\label{barPenRSurvive23}
\end{align}
For $(\bX_1,\bY) = (\bx,\by)\in\calH_n(B)$, the inner probability term at the r.h.s. of \eqref{barPenRSurvive23} can be represented as
\begin{align}
\Pr\ppp{\abs{\bX_2^T\by}\geq \abs{\bx^T\by}} &= 2\cdot \Pr\ppp{\bX_2^T\by\geq \abs{\bx^T\by}}\\
&\doteq \frac{\mathrm{Vol}\left(\mathcal{V}(\bx,\by) \cap \mathcal{T}_\varepsilon^n\right(P_X))}{\mathrm{Vol}(\mathcal{T}_\varepsilon^n(P_X))}\label{lowUppermar23}
\end{align}
where
\begin{align}
\mathcal{V}(\bx,\by)\triangleq\ppp{\bx'\in\mathbb{R}^n:\;\sum_{i=1}^nx_i'y_i\geq \abs{\sum_{i=1}^nx_iy_i}}.
\end{align}
Accordingly, using the same methods as in Appendix~\ref{app:1}, one can show that
\begin{align}
\mathrm{Vol}\left(\mathcal{V}(\bx,\by) \cap \mathcal{T}_\varepsilon^n\right(P_X))\doteq \exp\ppp{\frac{n}{2}\log\pp{2\pi eP_X\p{1-\frac{\abs{\sum_{i=1}^nx_iy_i}^2}{P_X\cdot\sum_{i=1}^ny_i^2}}}},
\end{align}
and
\begin{align}
\mathrm{Vol}(\mathcal{T}_\varepsilon^n(P_X))\doteq\exp\ppp{\frac{n}{2}\log\p{2\pi eP_X}}.
\end{align}
Thus, upon substitution in \eqref{lowUppermar23} and \eqref{barPenRSurvive23}, and using the LLN, we get that the probability of error converges to zero as $n\to\infty$ as long as,
\begin{align}
R&<-\frac{1}{2}\log\pp{1-\frac{h_0^2P_X}{\norm{\bh}_2^2P_X+\sigma^2}}\\
& =\frac{1}{2}\log\pp{1+\frac{h_0^2P_X}{(\norm{\bh}_2^2-h_0^2)\cdot P_X+\sigma^2}}.
\end{align}

\ifCLASSOPTIONcaptionsoff
  \newpage
\fi
\bibliographystyle{IEEEtran}
\bibliography{strings}

\begin{thebibliography}{10}
\providecommand{\url}[1]{#1}
\csname url@samestyle\endcsname
\providecommand{\newblock}{\relax}
\providecommand{\bibinfo}[2]{#2}
\providecommand{\BIBentrySTDinterwordspacing}{\spaceskip=0pt\relax}
\providecommand{\BIBentryALTinterwordstretchfactor}{4}
\providecommand{\BIBentryALTinterwordspacing}{\spaceskip=\fontdimen2\font plus
\BIBentryALTinterwordstretchfactor\fontdimen3\font minus
  \fontdimen4\font\relax}
\providecommand{\BIBforeignlanguage}[2]{{%
\expandafter\ifx\csname l@#1\endcsname\relax
\typeout{** WARNING: IEEEtran.bst: No hyphenation pattern has been}%
\typeout{** loaded for the language `#1'. Using the pattern for}%
\typeout{** the default language instead.}%
\else
\language=\csname l@#1\endcsname
\fi
#2}}
\providecommand{\BIBdecl}{\relax}
\BIBdecl

\bibitem{MerhavLapidoth}
N.~Merhav, G.~Kaplan, A.~Lapidoth, and S.~Shamai~(Shitz), ``On information
  rates for mismatched decoders,'' \emph{IEEE Trans. on Inf. Theory}, vol.~40,
  no.~5, pp. 1953--1967, Nov. 1994.

\bibitem{LapidothNarayan}
A.~Lapidoth and P.~Narayan, ``Reliable communication under channel
  uncertainty,'' \emph{IEEE Trans. on Inf. Theory}, vol.~44, no.~5, pp.
  2148--2177, Oct. 1998.

\bibitem{Ganti}
A.~Ganti, A.~Lapidoth, and E.~Telatar, ``Mismatched decoding revisited:
  {G}eneral alphabets, channels with memory, and the wide-band limit,''
  \emph{IEEE Trans. Inf. Theory}, vol.~46, no.~7, p. 2315–2328, Nov. 2000.

\bibitem{SomekhGene}
A.~Somekh-Baruch, ``A general formula for the mismatch capacity,'' \emph{IEEE
  Trans. on Inf. Theory}, vol.~61, no.~9, pp. 4554--4568, June 2015.

\bibitem{Stiglitz}
I.~G. Stiglitz, ``Coding for a class of unknown channels,'' \emph{IEEE Trans.
  on Inf. Theory}, vol. IT-12, pp. 189--195, Apr. 1966.

\bibitem{KaplanSh}
A.~Kaplan and S.~Shamai~(Shitz), ``Information rates of compound channels with
  application to antipodal signaling in a fading environment,'' \emph{A{E}U},
  vol.~47, no.~4, pp. 228--239, Nov. 1993.

\bibitem{abou_faycal}
I.~C. Abou-Faycal, ``An information theoretic study of reduced-complexity
  receivers for intersymbol interference channels,'' Ph.D. dissertation, MIT,
  2001.

\bibitem{Graph_decomposition}
I.~Csisz\'{a}r and J.~K{\"o}rner, ``Graph decomposition: A new key to coding
  theorems,'' \emph{IEEE Trans. on Inf. Theory}, vol. IT-27, pp. 5--12, Jan.
  1981.

\bibitem{Hui}
J.~Y.~N. Hui, \emph{Fundamental issues of multiple accessing}.\hskip 1em plus
  0.5em minus 0.4em\relax Ph.D. dissertation, Mass. Inst. Technol., Cambridge,
  MA, 1983.

\bibitem{ScarlettMis}
J.~Scarlett, A.~Martinez, and A.~Guill\'{e}n~i F\`{a}bregas, ``Mismatched
  decoding: Error exponents, second-order rates and saddlepoint
  approximations,'' \emph{IEEE Trans. on Inf. Theory}, vol.~60, no.~5, pp.
  2647--2666, May 2014.

\bibitem{Gallager}
R.~G. Gallager, \emph{Information Theory and Reliable Communication}.\hskip 1em
  plus 0.5em minus 0.4em\relax New York: Wiley, 1968.

\bibitem{CsisKro}
I.~Csisz\'{a}r and J.~K{\"o}rner, \emph{Information Theory: Coding Theorems for
  Discrete Memoryless Systems}.\hskip 1em plus 0.5em minus 0.4em\relax Academic
  Press, 1981.

\bibitem{AbbeMedard}
E.~Abbe, M.~Medard, S.~Meyn, and L.~Zheng, ``Finding the best mismatched
  detector for channel coding and hypothesis testing,'' in \emph{2007
  Information Theory and Applications Workshop}, Jan 2007, pp. 284--288.

\bibitem{Polyanskiy}
V.~Polyanskiy, V.~Poor, and S.~Verd\'u, ``Channel coding rate in the finite
  blocklength regime,'' \emph{IEEE Trans. on Inf. Theory}, vol.~56, no.~5, pp.
  2307--2359, May 2010.

\bibitem{Hayashi}
M.~Hayashi, ``Information spectrum approach to second-order coding rate in
  channel coding,'' \emph{IEEE Trans. on Inf. Theory}, vol.~55, no.~11, pp.
  4947--4966, Nov. 2009.

\bibitem{Jensen}
J.~L. Jensen, \emph{Saddlepoint Approximations}.\hskip 1em plus 0.5em minus
  0.4em\relax Oxford, U.K.: Oxford Univ. Press, 1995.

\bibitem{NarayanCsisz}
I.~Csisz\'{a}r and P.~Narayan, ``Channel capacity for a given decoding
  metric,'' \emph{IEEE Trans. on Inf. Theory}, vol.~41, no.~1, pp. 35--43, Jan.
  1995.

\bibitem{Bruijn}
N.~G. De~Bruijn, \emph{Asymptotic Methods in Analysis}.\hskip 1em plus 0.5em
  minus 0.4em\relax Dover Publications, Inc. New York, 1981.

\bibitem{NeriUni}
N.~Merhav, ``Universal decoding for memoryless gaussian channels with a
  deterministic interference,'' \emph{IEEE Trans. on Inf. Theory}, vol.~39,
  no.~4, pp. 1261--1269, July 1993.

\bibitem{Wasim3}
W.~Huleihel and N.~Merhav, ``Universal decoding for gaussian intersymbol
  interference channels,'' \emph{IEEE Trans. Inform. Theory}, vol.~61, no.~4,
  pp. 1606--161, Apr. 2015.

\bibitem{Merhavcodes}
N.~Merhav, ``Relations between random coding exponents and the statistical
  physics of random codes,'' \emph{IEEE Trans. on Inf. Theory}, vol.~55, no.~1,
  pp. 83--92, Jan. 2009.

\bibitem{MerhavSlep}
------, ``Erasure/list exponents for {S}lepian-{W}olf decoding,'' \emph{IEEE
  Trans. on Inf. Theory}, vol.~60, no.~8, pp. 4463--4471, Aug. 2014.

\bibitem{Merhavoptimalbin}
------, ``Exact random coding exponents of optimal bin index decoding,''
  \emph{IEEE Trans. on Inf. Theory}, vol.~60, no.~10, pp. 6024--6031, Oct.
  2014.

\bibitem{WasimInter}
W.~Huleihel and N.~Merhav, ``Exact random coding error exponents for the
  two-user interference channel,'' \emph{submitted to IEEE Trans. Inform.
  Theory}, Mar. 2015.

\bibitem{Csis2}
I.~Csisz\'ar, ``Linear codes for sources and source networks: error exponents,
  universal coding,'' \emph{IEEE Trans. on Inf. Theory}, vol. IT-28, no.~4, pp.
  585--592, July 1982.

\bibitem{ZivUni}
J.~Ziv, ``Universal decoding for finite-state channels,'' \emph{IEEE Trans. on
  Inf. Theory}, vol. IT-31, no.~4, pp. 453--460, July 1985.

\bibitem{LapZiv}
A.~Lapidoth and J.~Ziv, ``On the universality of the lz–based noisy channels
  decoding algorithm,'' \emph{IEEE Trans. on Inf. Theory}, vol.~44, no.~5, pp.
  1746--1755, Sep. 1998.

\bibitem{FerderLapidoth}
M.~Feder and A.~Lapidoth, ``Universal decoding for channels with memory,''
  \emph{IEEE Trans. on Inf. Theory}, vol.~44, no.~5, pp. 1726--1745, Sep. 1998.

\bibitem{merFeder}
M.~Feder and N.~Merhav, ``Universal composite hypothesis testing: a competitive
  minimax approach,'' \emph{IEEE Trans. on Inf. Theory special issue in memory
  of Aaron D. Wyner}, vol.~48, no.~6, pp. 1504--1517, June 2002.

\bibitem{Lomnitz}
\BIBentryALTinterwordspacing
Y.~Lomnitz and M.~Feder, ``Communication over individual channels – a general
  framework,'' Mar. 2012. [Online]. Available: \url{arXiv:1023.1406v1}
\BIBentrySTDinterwordspacing

\bibitem{Lomnitz2}
\BIBentryALTinterwordspacing
------, ``Universal communication over modulo–additive channels with an
  individual noise sequence,'' May. 2012. [Online]. Available:
  \url{arXiv:1012.2751v2}
\BIBentrySTDinterwordspacing

\bibitem{Misra}
\BIBentryALTinterwordspacing
V.~Misra and T.~Weissman, ``The porosity of additive noise sequences,'' May.
  2012. [Online]. Available: \url{arXiv:1025.6974v1}
\BIBentrySTDinterwordspacing

\bibitem{Shayevitz}
O.~Shayevitz and M.~Feder, ``Communicating using feedback over a binary channel
  with arbitrary noise sequence,'' in \emph{Proc. ISIT 2005}, Sep. 2005, pp.
  1516--1520.

\bibitem{Shayevitz2}
------, ``Universal decoding for frequency-selective fading channels,''
  \emph{IEEE Trans. on Inf. Theory}, vol.~51, no.~8, pp. 2770--2790, Aug. 2005.

\bibitem{Shayevitz3}
------, ``Universal decoding for frequency-selective fading channels,''
  \emph{IEEE Trans. on Inf. Theory}, vol.~51, no.~8, pp. 2770--2790, Aug. 2005.

\bibitem{UniNeri2}
N.~Merhav, ``Universal decoding for arbitrary channels relative to a given
  class of decoding metrics,'' \emph{IEEE Trans. on Inf. Theory}, vol.~59,
  no.~9, pp. 5566--576, Sep. 2013.

\bibitem{HIRTMASSEY}
W.~Hirt and J.~L. Massey, ``Capacity of the discrete-time gaussian channel with
  intersymbol interference,'' \emph{IEEE Transactions on Information Theory},
  vol.~34, no.~3, pp. 38--38, 1988.

\bibitem{MerhavArbitrary}
N.~Merhav, ``Universal decoding for arbitrary channels relative to a given
  class of decoding metrics,'' \emph{IEEE Transactions on Information Theory},
  vol.~59, no.~9, pp. 5566--5576, Sept 2013.

\bibitem{Shulman}
N.~Shulman, \emph{Communication over an Unknown Channel via Common
  Broadcasting}.\hskip 1em plus 0.5em minus 0.4em\relax Ph.D. dissertation,
  Department of Electrical Engineering – Systems, Tel Aviv University, July
  2003.

\bibitem{Folland}
G.~B. Folland, \emph{Real Analysis}.\hskip 1em plus 0.5em minus 0.4em\relax
  John Wiley \& Sons, 1984.

\bibitem{Szego}
U.~Grenander and G.~Szego, \emph{Toeplitz Forms and Their Applications}.\hskip
  1em plus 0.5em minus 0.4em\relax University of Calif. Press, Berkeley and Los
  Angeles, 1958.

\bibitem{TimeSeriesBook}
S.~Ihara, \emph{Information Theory for Continuous Systems}.\hskip 1em plus
  0.5em minus 0.4em\relax Series on Probability \& Statistics, World Scientific
  Pub Co Inc, 1993.

\bibitem{ChengVerdu}
R.~S. Cheng and S.~Verd{\'u}, ``Gaussian multiaccess channels with isi:
  Capacity region and multiuser water-filling,'' \emph{IEEE Transactions on
  Information Theory}, vol.~39, no.~3, pp. 773--785, 1993.

\bibitem{scarletNew}
J.~Scarlett, A.~Martinez, and A.~Guill\'{e}n~i F\`{a}bregas, ``Multiuser coding
  techniques for mismatched decoding,'' \emph{IEEE Trans. on Inf. Theory},
  vol.~62, no.~7, pp. 3950--3970, July 2016.

\bibitem{HuleihelMerhav}
W.~Huleihel and N.~Merhav, ``Random coding error exponents for the two-user
  interference channel,'' \emph{IEEE Trans. on Inf. Theory}, vol.~63, no.~2,
  pp. 1019--1042, Feb. 2017.

\bibitem{NeriMono}
N.~Merhav, ``Statistical physics and information theory,'' \emph{Foundations
  and Trends in Communications and Information Theory}, vol.~6, no. 1-2, pp.
  1--212, Dec. 2010.

\bibitem{remmert2012theory}
R.~Remmert, \emph{Theory of complex functions}.\hskip 1em plus 0.5em minus
  0.4em\relax Springer Science \& Business Media, 2012, vol. 122.

\end{thebibliography}
\end{document}